\documentclass{ws-ijmpe}
\usepackage[super,compress]{cite}

\usepackage{color} 
\usepackage{soul} 

\newcommand{\lsc}{\lambda_{sc}}

\begin{document}

\markboth{S.~A.~Coon and M.~K.~G.~Kruse}{Properties of infrared extrapolations in a harmonic oscillator basis}

\catchline{}{}{}{}{}

\title{\bf Properties of infrared extrapolations in a harmonic oscillator basis}
\author{Sidney A. Coon}
\address{Department of Physics, University of Arizona\\ Tucson, Arizona USA \\
coon@physics.arizona.edu}

\author{Michael K. G. Kruse}

\address{ Lawrence Livermore National Laboratory, P.O. Box 808, L-414\\
Livermore, California 94551, USA\\
kruse9@llnl.gov}

\maketitle

\begin{history}
\received{Day Month Year}
\revised{Day Month Year}
\end{history}

\begin{abstract}

 The success and utility of effective field theory (EFT) in explaining the structure and reactions of few-nucleon systems has prompted the initiation of EFT-inspired extrapolations to larger model spaces in {\em ab initio} methods such as the  no-core shell model (NCSM). In this contribution, 
we review and continue our studies of infrared (ir) and ultraviolet (uv) regulators of   NCSM calculations in which the input is phenomenological   $NN$ and $NNN$ interactions fitted to data.  We extend our previous findings  that 
 an extrapolation in the ir cutoff with the uv cutoff above the intrinsic uv scale of the interaction is quite successful, not only for the eigenstates of the Hamiltonian but also for  expectation values of operators, such as $r^2$, considered long range.  The latter results are obtained with Hamiltonians transformed by the similarity renormalization group (SRG) evolution.  On the other hand, a possible extrapolation  of ground state energies in the uv cutoff when the ir cutoff is below the intrinsic ir scale is not robust and does not agree with the ir extrapolation of the same data or with independent calculations using other methods.
 \end{abstract}


\ccode{PACS numbers:21.60.De,  21.45.-v,  13.75.Cs,  21.30.-x}


\section{Introduction}
Variational calculations based upon a harmonic oscillator (HO) basis expansion have a long history in nuclear structure physics.   
The traditional shell-model calculation, often for nuclear spectroscopy, involves wave functions which are linear combinations of Slater determinants.   Each Slater determinant corresponds to a configuration of $A$
fermions distributed over  a set of single-particle states, as defined by the basis truncation scheme. In the traditional shell model the valence nucleons of an open shell are only allowed to occupy the single-particle states of the open-shell whereas the core (a closed shell) is energetically frozen. In modern {\em ab-initio} calculations the concept of a core is abandoned and all $A$ nucleons can be distributed over the set of single-particle states.  If we take any complete set of orthonormal single-particle wave functions and consider all possible $A$-particle Slater determinants that can be formed from them, then these  wave functions form a complete orthonormal set of wave functions spanning the $A$-particle Hilbert space.  Eigenfunctions of the three-dimensional harmonic oscillator (HO) are a popular choice. 
 If one  views a shell-model calculation as a variational calculation, expanding the configuration space merely serves to improve the trial wave function \cite{Irvine}.  Another program  uses the HO eigenfunctions as a basis of a finite linear expansion to make a straightforward {\itshape ab initio}  variational calculation of ground state properties of light nuclei \cite{Moshinsky}.    The trial functions take the form of a finite linear expansion in a set of known functions 
 \[\Psi_T = \sum_{\nu}^{\mathcal{N}}a_\nu^{({\mathcal{N}})}h_\nu  \]
 where $a_\nu^{({\mathcal{N}})}$ are the parameters to be varied and  $h_\nu$ are many-body states based on a summation over products of HO functions. The expansion coefficients 
depend on the upper limit (such as an $\mathcal{N}$ defined in terms of total oscillator quanta) and are obtained by minimizing the expectation value of the  Hamiltonian in this basis. 
Treating the coefficients $a_\nu^{({\mathcal{N}})}$ as variational parameters in the Rayleigh quotient,  
one performs the variation by diagonalizing the many-body Hamiltonian in this basis.
This is an eigenvalue problem so the minimum with respect to the vector of expansion coefficients always exists and one obtains a bound on the lowest eigenvalue (and indeed on the higher eigenvalues representing the excited states).~\cite{Don33}  The basis functions can also depend upon a  parameter (such as the harmonic oscillator energy $\hbar\omega$ which sets a scale)  that then becomes a  non-linear variational parameter additional to the linear expansion coefficients. 

Theorems based upon functional analysis established the asymptotic convergence rate of these   calculations as a function of the counting number ($\mathcal{N}$) which characterizes the size of the expansion basis (or model space) \cite{Delves72, Sch72}.  In addition, a recent discussion of the configuration-interaction method in a HO basis analyzes convergence for many-electron systems trapped in a harmonic oscillator (a typical model for a quantum dot) \cite{Kvaal}. The convergence rates  of the functional analysis  theorems (inverse power laws in $\mathcal{N}$ for ``non smooth" potentials such as Yukawas with strong short range correlations and exponential in  $\mathcal{N}$ for ``smooth" potentials such as gaussians) were demonstrated numerically in Ref.~\refcite{Delves72} for the HO expansion and in 
Ref.~\refcite{Fabre} for the  analogous expansion in hyperspherical harmonics.  These  convergence   theorems underlie extrapolations to the ``infinite" basis in few-nucleon studies \cite{JLS70} and in ``{\itshape ab initio}" ``no-core shell model" (NCSM) calculations of $s$- and $p$-shell nuclei reviewed in Ref. \refcite{NQSB}.

 However, the HO expansion basis has  an intrinsic scale parameter $\hbar\omega$ which does not naturally fit into an extrapolation scheme based upon $\mathcal{N}$ as discussed in  
  Refs.~\citen{Delves72, Sch72, Kvaal}.  Indeed the model spaces of these NCSM approaches are characterized by the pair ($\mathcal{N},\hbar\omega$).   Here the basis truncation parameter $\mathcal{N}$ and the HO energy parameter $\hbar\omega$ are variational parameters \cite{NQSB,NavCau04, Maris09}, provided the two-body interaction is ``soft" enough.  
 Convergence has been discussed, in practice,  with an emphasis on obtaining those parameters which appear linearly in the trial function (i.e. convergence with $\mathcal{N}$). In an early example, $\hbar\omega$ is simply fixed at a value which gives the fastest convergence in $\mathcal{N}$ \cite{JLS70}. Later,  for each $\mathcal{N}$ the  non-linear parameter $\hbar\omega$ is varied to obtain the minimal energy  \cite{CP87,NavCau04} for a fixed  $\mathcal{N}$ and then the convergence with $\mathcal{N}$ is examined at that fixed value of $\hbar\omega$.     Other extrapolation schemes have been proposed and used \cite{Maris09,Forssen2}.

  \begin{figure}
    \centering
     \includegraphics[width=0.90\textwidth]{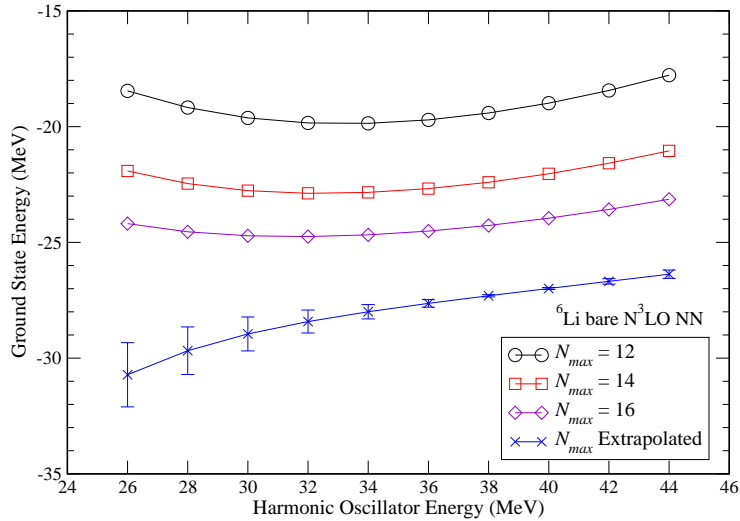}
\caption{(Color online) Harmonic-oscillator energy dependence of the $^6$Li ground-state energy obtained in model spaces labeled by $N_{max} = 12-16$
 using the bare  N$^3$LO $NN$ potential.  The extrapolated result in $N_{max}$ at each HO energy is shown along with estimated error bars.  For details of the extrapolation see the text.}
\label{fig:1}
\end{figure}

 Figure 1 demonstrates the unsatisfactory aspect of the traditional extrapolation scheme which relies primarily on convergence with $\mathcal{N}$ at a fixed  value of $\hbar\omega$.  In this example, 
  $\mathcal{N} = N_{max}$,  the maximum of the {\em total} HO quanta shared by all nucleons above the lowest  HO configuration of the nucleus $^6$Li.  This  truncation, a characteristic feature of the NCSM,  is  at the level of total energy quanta (``total-energy-cut space"), which is different  from the configuration-interaction calculations  used in atomic and molecular problems, that are often truncated at the single-particle level (``single-particle-cut space").  From the figure, it is clear that the calculation has not yet converged (in $\mathcal{N}$) at the largest $\mathcal{N}$ available. The extrapolation scheme of  Refs.~\citen{Delves72, Sch72, Kvaal} has been used in Ref.~\refcite{NavCau04} for  this nucleus subject to the Idaho N$^3$LO  $NN$ interaction \cite{IdahoN3LO}.
The extrapolation is performed by a fit of an exponential plus a constant to each set of results at fixed  $\hbar\omega$.  That is, we fit the ground state energy with three adjustable parameters using the relation

 \begin {equation}  E_{gs}(\hbar\omega,N_{max}) = a_{\hbar\omega} \exp(-b_{\hbar\omega}N_{max}) + E_{gs}( \hbar\omega   ,N_{max}=\infty)  .   \label{eq:traditional}
 \end{equation}

The extrapolation is done twice,  for  $N_{max}=10-16$ and  $N_{max}=8-14$,  the average is denoted by the $\times$ and half the difference of the extrapolations is considered the uncertainty denoted by error bars.  Notice that the fixed $N_{max}$ curves do not appear to depend strongly upon the HO energy 
 $\hbar\omega$ for the largest  $N_{max}$ and the minimum of this curve is often chosen as that point at which one makes the extrapolation in $N_{max}$ \cite{CP87,Maris09,NavCau04}.  However, as shown in Figure 1, the extrapolated ground state energy does depend rather strongly upon $\hbar\omega$, everywhere in the range displayed. This strong dependence is true for the values $\hbar\omega = 32-34$ MeV at the minimum of the lowest fixed $N_{max}$ curve, values which would be the traditional choice for the extrapolated result.  In reality, since both the basis truncation $\mathcal{N}$  and $\hbar\omega$ are variational parameters and the dependence upon $\hbar\omega$ remains in the extrapolated values of Figure 1, convergence has not been reached in  both variational parameters. 
 The same pattern of a strong $\hbar\omega$ dependence of the extrapolated result for the calculated nucleus $^7$Li subject to the JISP16 $NN$ interaction \cite{Shirokov07} is observed in Figure 1 of Ref.~\refcite{Maris_NTSE-2013} .  The traditional extrapolation to an infinite basis in  the  context of the no-core Monte Carlo shell model \cite{Abe_NTSE-2014}, which employs a truncation at the single-particle level rather than $N_{max}$, also shows a strong $\hbar\omega$ dependence of the extrapolated ground state energies of $^4$He, $^8$Be, $^{12}$C and $^{16}$O .
 We  conclude that the ground state energy cannot be determined from the traditional extrapolation in these cases.  Another, less pessimistic interpretation, is that of Ref.~\refcite{Abe_NTSE-2014} which shows in its Table 1 a large error with the traditional extrapolation, due to the  $\hbar\omega$ dependence, even for the apparently converged $^4$He.

It is the purpose of this study to review and continue an investigation of the   extrapolation tools introduced in Ref.~\refcite{Coon2012}  which use  $\mathcal{N}$ and $\hbar\omega$ on an equal footing.  These tools are based upon 
the pair of ultraviolet (uv) and infrared (ir) cutoffs (each a function of both $\mathcal{N}$ and $\hbar\omega$) of the model space.  Ultraviolet and infrared regulators were first introduced to the NCSM by Ref.~\refcite{EFTNCSM} in the context of an effective field theory (EFT) approach.  Let us briefly mention the salient aspects of EFT.~\cite{Birareview}  In a field theory one never has access to the ÒfullÓ Hilbert space. Experiments only probe a region of momenta. Nature is quantum mechanical. So to develop a theory for such a region we must pose a model space. For smallest errors the model space should be as big, if possible, as the region one is interested in. The parameter of the projection operator P into the model space must have a dimension. Call the parameter $\Lambda$, the ultraviolet cutoff and take it to be a momentum.  The model space can be arbitrary but observables calculated within it cannot.
The Hamiltonian operator of the model space must depend on $\Lambda$ in such a way that observables at momenta  $Q\ll \Lambda$ are independent of how P is chosen, and in particular, independent of $\Lambda$. In the NCSM case, the HO basis results in a second parameter $\lambda$, an infrared cutoff in addition to $\Lambda$, so that observables at momenta $Q\gg \lambda$ should be independent of  $\lambda$. That is, the values of   $\Lambda$ and  $\lambda$ control the size of the model space and the projection operators P($\Lambda$) and P($\lambda$) define the boundaries of the model space.  
In the EFT approach to the NCSM, a Hamiltonian is always constructed within  this truncated model space according to the symmetries of the underlying theory, making use of power counting to limit the number of interactions included in the calculations. Hence, physical terms not explicitly included in the calculation are treated on the same footing as the truncation to a finite model space.  For a recent review of this program applied to the NCSM see  Ref.~\citen{Ionel_Jimmy}.  

In contrast, the program started by us in Ref.~\refcite{Coon2012} and continued in this paper uses   EFT concepts to motivate and guide an extrapolation to the infinite basis limit  of those NCSM calculations which utilize ÒrealisticÓ nuclear interactions fit to data, not in a clearly defined model space as in Ref. \refcite{EFTNCSM}, but in free space.  
A parallel program has been carried out in Refs.~\citen{FHP,later1,later2,later3,later4}.   This latter program 
derived  an  ir extrapolation formula of the form of our equation~\ref{eq:irextrap} in Section  3   for a single-particle in three dimensions (two-body bound state in the center of mass).   A uv extrapolation formula was not derived in Ref.~\refcite{FHP} but obtained empirically. These extrapolation formulae, with the   ir regulator parameter interpreted in terms of a single-particle separation energy,  were applied in Ref.~\refcite{FHP} to coupled-cluster  calculations of the $A >  2$ nuclei $^{16}$O  and $^6$He .  The coupled-cluster method is an example of a ``single-particle-cut space" and is reviewed in Ref.~\refcite{ccreview}. 
Other workers applied the extrapolations advocated in  Refs.~\citen{FHP,later1} to NCSM calculations of $^6$He in Ref.~\refcite{Forssen6He}  and of  other  $p$-shell nuclei in Ref.~\refcite{Livermore}.

 The early {\itshape ab initio} calculations, both of the no-core shell model in which all nucleons are active \cite{Irvine}  and of the Moshinsky program  \cite{Moshinsky} attempted to overcome the challenges posed by ``non-smooth" two-body potentials  by including Jastrow type two-body correlations in the trial wave function.   Nowadays, the $NN$ potentials are tamed by either {\it i}) unitary transformations within the model space \cite{OSL} or {\it ii})  in free space by  the similarity renormalization group (SRG) evolution \cite{BFP,Bognerreview,Jurgenson, FHeblerreview} or the $V_{low\: k}$ truncation \cite{debate}.  In all three cases, this procedure generates effective many-body interactions in the new Hamiltonian.  Neglecting these in  {\it i}) destroys the variational aspect of the calculation (and changes the physics contained in the calculation, of course).  Neglecting these in {\it ii}) destroys the unitary nature of the transformation (and changes the physics contained in the calculation, of course).  We  retain the variational nature of our NCSM investigation  by choosing a realistic smooth nucleon-nucleon ($NN$) interaction Idaho N$^3$LO \cite{IdahoN3LO}  which has been used previously without renormalization within the model space for  light nuclei ($A \leq$ 6) \cite{NavCau04}.   
 The Idaho N$^3$LO potential is softer than those used before, with high-momentum components that are heavily reduced by the regulator (``super-Gaussian falloff in momentum space")  as compared to earlier realistic $NN$ potentials which had a strongly repulsive core.  Alternatively, in coordinate space, the contact interaction and the Yukawa singularity at the origin are regulated away so that this potential would be considered ``smooth" by Delves and 
Schneider and the convergence in $\mathcal{N}$ would be expected to be exponential \cite{Delves72, Sch72}.  Even without the construction of an effective interaction, convergence with the Idaho N$^3$LO $NN$ potential is exponential in $\mathcal{N}$ at fixed $\hbar\omega$, as numerous studies have shown \cite{NavCau04,Jurgenson}.     We also illustrate concepts with a second $NN$ interaction; JISP16 \cite{Shirokov07} which is a nonlocal separable potential whose form factors are HO wave functions.  By construction, it is also a ``smooth" potential and  it also has  an exponential convergence  in $\mathcal{N}$ at fixed $\hbar\omega$ as demonstrated in many  variational studies.\cite{Maris09, Cockrell12, Maris13}

 We  refer the reader to a comprehensive  review article \cite{NQSB} on the no-core shell model (NCSM) for  details  of  HO bases and to a later review  \cite{BNV} for more explication and references to the recent literature.  We used for our calculations 
  technology developed and/or adapted for NCSM, such as the shell model code ANTOINE \cite{ANTOINE},  the {\it manyeff} code from Ref. \refcite {NKB},
and the No-Core Shell Model Slater Determinant Code \cite{NCSMSD} and we quote calculations for the JISP16 potential which were made with the  parallel-processor code ``Many-Fermion Dynamics --- nuclear" (MFDn) \cite{Vary92_MFDn}.

The paper is organized as follows.  In section 2   we review the  uv and ir cutoffs introduced in Ref. \refcite{EFTNCSM} in the context of an EFT framework and discussed in the context of NCSM extrapolations in Ref. \refcite{Coon2012}.  We  show the running of ground state energies with the cutoffs and relate the cutoffs to scales intrinsic to the $NN$ interaction in Section 3.   Examples of convergence and extrapolation with these regulator functions are displayed in Section 4.  Section 5 contains a short summary.

\section{Ultraviolet and infrared cutoffs inherent to the finite HO basis}

Inspired by EFT, one uses a truncation parameter  $\mathcal{N}$ which refers, not to the many-body system, but to the properties of the HO single-particle states.  The many-body truncation parameter  $N_{max}$ is the maximum number of oscillator quanta
shared by all nucleons above the lowest HO configuration for the
chosen nucleus.  One unit of oscillator quanta is one unit of the quantity $(2n+l)$
where $n$ is the principle quantum number and $l$ is the orbital
angular quantum number.
If the highest HO single-particle state of this lowest HO
configuration has $N_0$ HO quanta, then $N_{max}+N_0=N$ identifies the
highest HO single-particle states that can be occupied within this
many-body basis.  Since
$N_{max}$ is the maximum of the {\em total} HO quanta above the
minimal HO configuration, we can have at most one nucleon in such a
highest HO single-particle state with $N$ quanta.  Note that $N_{max}$ characterizes the many-body basis space, whereas $N$ is a label of the corresponding single particle space.  Let us illustrate this distinction with two  examples.  $^6$He is an open shell nucleus with $N_0=1$  since both valence neutrons
occupy the $0p$  shell in the lowest energy many-body configuration.
Assigning a single neutron  the entire $N_{max}=4$ quanta means that the
highest occupied single-particle state is in the $N=5$ shell. On the other
hand, the highest occupied orbital of the closed
s-shell nucleus $^4$He has $N_0=0$  so that $N=N_{max}$.

We begin the transition to uv and ir regulators by thinking of the finite single-particle basis space defined by $N$
and $\hbar\omega$ as a model space characterized by two momenta associated with the basis functions themselves.
We follow Ref.~\refcite{EFTNCSM} and define $\Lambda=\sqrt{m_N(N+3/2)\hbar\omega}$ as the 
momentum (in units of MeV/$c$) associated with the energy of the highest HO level.  The nucleon mass is taken to be the average mass  $m_N=938.92$ MeV. To arrive at this definition one applies the virial theorem to this highest HO level  to establish  kinetic energy as one half the total energy ({\it i.e.}, $(N+3/2)\hbar\omega\:$)  and solves the non-relativistic dispersion relation for $\Lambda$. 
 Thus, the usual definition of an ultraviolet cutoff $\Lambda$ in the continuum has been extended to discrete HO states. It is then quite natural to interpret the behavior of the variational energy of the system with addition of more basis states as the behavior   of this observable with the running of the ultraviolet cutoff $\Lambda$.    
  Because the energy levels of a particle in a HO potential are quantized in units of $\hbar\omega$,  the  momentum difference between single-particle orbitals  is 
$\lambda=\sqrt{m_N\hbar\omega}$ and that has been taken to be an infrared cutoff \cite{EFTNCSM}.  That is, the postulated   low-momentum cutoff $\lambda=\hbar/b$  where $b=\sqrt{\frac{\hbar}{m_N\omega}}$ plays the role of a characteristic  length of the HO potential and basis functions.  
In Ref.~\refcite{EFTNCSM} the influence of the infrared cutoff is removed by extrapolating to the continuum limit, where  $ \hbar\omega\rightarrow 0$    with $N\rightarrow\infty$ so that $\Lambda$ is fixed.  Clearly, one cannot achieve both the ultraviolet limit and the infrared limit by taking $\hbar\omega$ to zero  in a fixed-$N$ model space as this procedure takes the ultraviolet cutoff to zero.

Other studies define the ir cutoff as  the infrared momentum which corresponds to the maximal
radial extent  needed to encompass the many-body system we are attempting to describe by the finite basis space (or model space).  These studies find it natural to define the ir cutoff by $\lambda_{sc}=\sqrt{(m_N\hbar\omega)/(N+3/2)}$ \cite {Jurgenson,Papenbrock}.  Note that $\lambda_{sc}$ is the inverse of the root-mean-square ({\it  rms}) radius of the highest single-particle state in the basis; $\langle r^2\rangle^{1/2}=b
\sqrt{N+3/2} $.   We distinguish the two definitions by  denoting the (historically) first  definition by $\lambda$ and the second definition by $\lambda_{sc}$ because of its  scaling properties demonstrated in the next Section.

 The calculated energies of a many-body system in the truncated model space will differ from those calculated as the basis size increases without limit ($N\rightarrow\infty$).  This is because the system is in effect confined within a finite (coordinate space) volume characterized by the finite value of $b$ intrinsic to the HO basis. The ``walls"  of the volume confining the interacting system spread apart  and the volume increases to the infinite limit  as $\lambda\rightarrow 0$ and $b\rightarrow\infty$ with $\Lambda$ held fixed.   
   The development of ir extrapolation formulae  in Ref.~\refcite{FHP} {\it et sequentes} is based upon the explicit  characterization of the finite confining coordinate space volume as a sphere with a hard boundary wall\cite{PineLee} and a radius related to $1/\lambda_{sc}$.   These energy level shifts in a large enclosure have long been studied since the first expositions \cite{FukudaNewton,PineLee}.  In condensed matter physics the relation between phase shifts and energy level shifts  is known as Fumi's theorem \cite{Mahan} and has found a recent application in lattice quantum chromodynamics (LQCD) calculations of nuclear matter with hyperon components \cite{NPLQCD}.  LQCD calculations are necessarily performed in a finite Euclidean spacetime. As a result, it is necessary to construct a formalism that maps the finite-volume observables determined via LQCD to the infinite-volume quantities of interest.  This formalism started with the work of L\"{u}scher \cite{Luscher} and is currently being extended extensively \cite{Raul}.

  Recently an EFT calculation of a triton in a cubic box allowed the edge lengths to become large so that the associated  ir cutoff due to momentum quantization in the box approaches zero \cite{Kreuzer}.  There it was shown that as long as the infrared cutoff was small compared to the ultraviolet momentum cutoff appearing in the ``pionless" EFT, the ultraviolet behavior of the triton amplitudes was unaffected by the finite volume.  More importantly, from our point of view of desiring extrapolation guidance, this result means that calculations in a finite volume can confidently be applied to the infinite volume (or complete model space) limit.  Similar conclusions can be drawn from the ongoing studies of systems of two and three nucleons trapped in a HO potential  \cite{moreEFT}. The model space interactions are from pionless EFT  and the infrared cutoff ($\lambda=\sqrt{m_N\hbar\omega}$) is taken to zero to remove the trapping potential
 ; see the review in Ref.~\refcite{Ionel_Jimmy}.

 \section{Running of variational energies with cutoffs and establishment of intrinsic regulator scales}

We display in the next three figures the running of the ground-state eigenvalue of a light nucleus  on the truncated HO basis by holding one cutoff of ($\Lambda,\lambda_{ir}$) fixed and letting the other vary.  In Figure 2 and the following figures,  $\vert\Delta E/E\vert $ is defined as $\vert (E(\Lambda,\lambda_{ir}) - E)/E\vert$ where $E$ reflects a consensus ground-state energy from benchmark calculations with this $NN$ potential,  this nucleus, and different few-body methods.  In NCSM calculations  the numerical accuracy of an  $E(\Lambda,\lambda_{ir})$ is  about a  keV in $A\leq 6$ nuclei for the values of  ($\Lambda,\lambda_{ir}$) considered in this investigation. The benchmark results  of $E$ for the  Idaho N$^3$LO interaction are summarized in,  for example,  Tables 1 and 2 of Ref.~\refcite{NQSB} and in Table 2 of Ref.~\refcite{BNV}. 
Benchmark results for the JISP16 potential are in Table 2 of Ref.~\refcite{EIHH} which shows excellent agreement between NCSM and the effective interaction hyperspherical harmonic expansion method and in Figure  2  of Ref.~\refcite{Abe} which displays similar consistency between NCSM and the no-core Monte Carlo shell model.

  \begin{figure}[!t]
    \centering
     \includegraphics[width=0.90\textwidth]{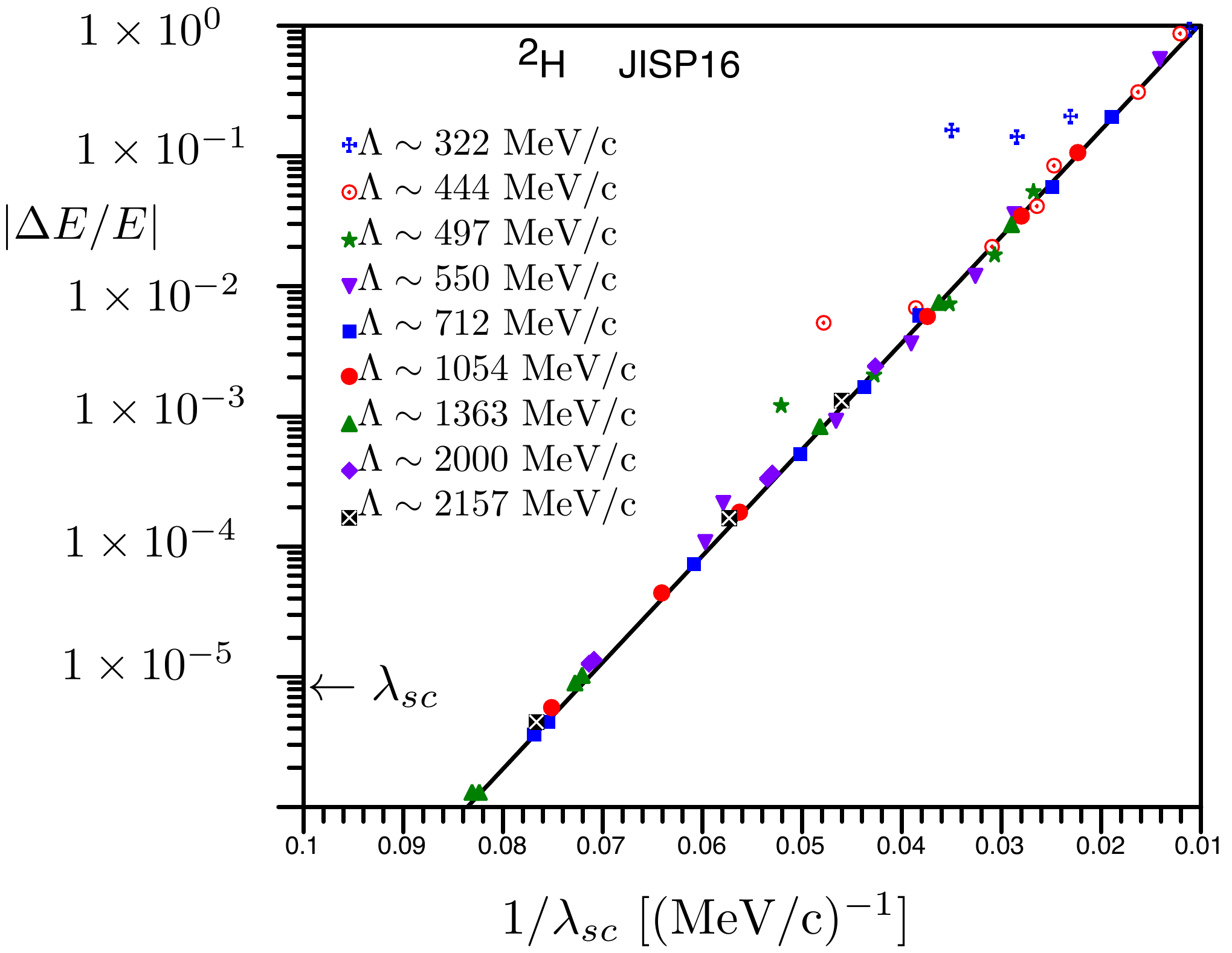}
\caption{(Color online) Dependence of the ground-state energy of $^2$H  upon the ir momentum cutoff  $\lambda_{sc}=\sqrt{(m_N\hbar\omega)/(N+3/2)}$  for fixed  $\Lambda=\sqrt{m_N(N+3/2)\hbar\omega}$.  
The curve is an exponential fit to the calculated points with $\Lambda\sim 712$ MeV. $\lsc$ decreases in magnitude from right to left as indicated by the arrow.} 
\label{fig:2}
\end{figure}

We choose for Figs. 2-4 the nucleus $^2$H, where exact results are known, to illustrate that the running of its ground state energy, already documented for $^3$H in Ref.~\refcite{Coon2012}, also holds in a case where the input values of ($N, \hbar\omega$)  are those of typical NCSM calculations.   
In Figure 2 we hold fixed the uv cutoff of ($\Lambda,\lambda_{ir}$) to display the running of  $\vert\Delta E/E\vert$ upon the suggested ir cutoff $\lambda_{sc}$.  Results with the JISP16 potential were obtained\cite{Maris09} on a mesh of integer 
($N,\hbar\omega$) so that the values of $\Lambda$ are not strictly fixed, but each point plotted corresponds to a value of $\Lambda$ constant to within  2\%-5\% of the central value indicated. For fixed $\lambda_{sc}$, a larger $\Lambda$ implies a smaller  $\vert\Delta E/E\vert$ since more of the uv region is included in the calculation.  For $\Lambda$ smaller than $\sim 500$ MeV, $\vert\Delta E/E\vert$ does not go to zero as the ir cutoff is lowered and more of the infrared region is included in the calculation.  This behavior suggests that $\vert\Delta E/E\vert$ does not go to zero unless $\Lambda\geq\Lambda^{NN}$, where $\Lambda^{NN}$ is some uv regulator scale of the $NN$ interaction itself.  From this figure one estimates  $\Lambda^{NN}\sim$ 500-550 MeV/$c$ for the JISP16 interaction.  For $\Lambda < \Lambda^{NN}$ there will be missing contributions 
  so ``plateaus" develop as $\lambda_{ir}\rightarrow 0$,  revealing this missing contribution to $\vert\Delta E/E\vert$.  The ``plateaus" that we do see  are not flat as $\lambda_{ir}\rightarrow 0$ and, indeed, rise significantly with decreasing $\Lambda <\Lambda^{NN}$.  This suggests that corrections are needed to  $\Lambda$ and $\lambda_{ir}$,  perhaps in the form of higher order terms in $\lambda_{ir}/\Lambda$; a subject for further study.
  
 Around  $\Lambda\sim 500$ MeV/$c$ and above the plot of $\vert\Delta E/E\vert$ versus $\lambda_{sc}$ in Figure 2  begins to suggest a universal pattern, especially at large $\lambda_{sc}$.
For $\Lambda\sim 700$ MeV/$c$ and above the pattern defines a universal curve for all values of  $\lambda_{sc}$.  This is  the region  where $\Lambda\geq\Lambda^{NN}$indicating that nearly all of the ultraviolet physics set by the potential has been captured.   
The universal curve can be fit by the equation $\vert\Delta E/E\vert   = {\cal A} \exp(-{\cal B}/ \lambda_{sc})$ which suggests immediately that  $\lambda_{sc}$ could be used for extrapolation to the ir limit ($\lambda_{sc} \rightarrow 0$), provided that $\Lambda$ is kept large enough to capture the uv region of the calculation, {\textit {i.e.}}  $\Lambda\geq\Lambda^{NN}$.    Figure 2 is also the motivation for our appellation $\lambda_{sc}$, which we read as ``lambda scaling", since  this figure exhibits the attractive scaling properties of this regulator.  

  \begin{figure}[ht]
       \centering
        \includegraphics[width=0.90\textwidth]{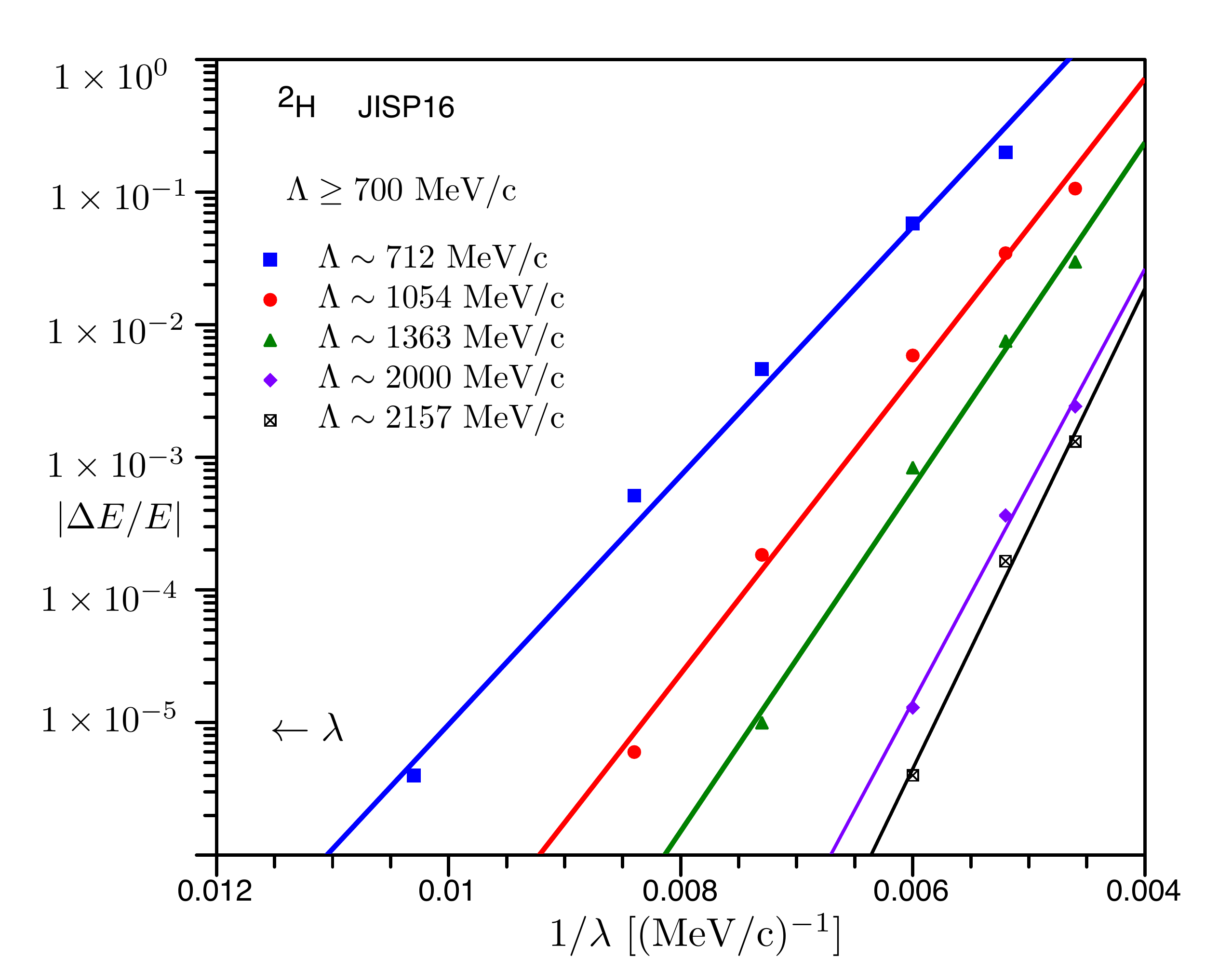}
\caption{(Color online) Dependence of the ground-state energy of $^2$H  upon the ir momentum cutoff  $\lambda=\sqrt{(m_N\hbar\omega)}$  for fixed  $\Lambda=\sqrt{m_N(N+3/2)\hbar\omega}$.  The curves are exponential fits to the calculated points. $\lambda$ decreases in magnitude from right to left as indicated by the arrow.}  
\label{fig:3}
\end{figure}

The first suggested  ir cutoff  $\lambda=\sqrt{m_N\hbar\omega}$, corresponding to the non-zero energy spacing between HO levels,  gives not a universal curve for $\Lambda\geq\Lambda^{NN}$  but instead a set of curves fit by $\vert\Delta E/E\vert   = {\cal A} \exp(-B(\Lambda)/ \lambda)$ (see Figure 3).  That is, the value of $B$ is  constant for a given  $\Lambda$ but that constant is not  independent of the value of the uv cutoff $\Lambda$, as it should be in an EFT framework.  One can remove the dependence of $B$ upon $\Lambda$ to a large extent by noting that $\lambda =\sqrt{\Lambda\lambda_{sc}}$ so that $\exp(-B/\lambda)$ becomes $\exp\left(\frac{-B/\sqrt{\Lambda}}{\sqrt{\lambda_{sc}}}\right)$.   This multiplier of $1/\sqrt{\lambda_{sc}}$  is constant to within a few per cent.  This trivial manipulation demonstrates that the ir regulator which is independent of the uv cutoff is a function of $\lambda_{sc}$ and not $\lambda$.  The point is not that the ir cutoff $\lambda$ cannot be used to remove ir effects by extrapolating it to zero.   Indeed, any momentum cutoff $\lambda_{sc} \leq \lambda_{ir} \leq \Lambda$ will remove ir artifacts, but the ir cutoff which is independent of the uv cutoff is some function of $\lambda_{sc}$.   One does not need to decrease an ir cutoff below that of $\lambda_{sc}$ to remove ir effects.

By ``remove ir artifacts", we mean that any given NCSM matrix diagonalization is performed in a model space defined by $(N,\hbar\omega)$ or, in an EFT type characterization, by $(\Lambda,\lambda_{ir})$.  As $N$ is limited to a finite value, 
this model space contains unwanted uv and ir effects (``artifacts").  As the physics should not depend upon the choice of the model space, these artifacts are removed from a sequence of results by taking $\lambda_{ir}$ to zero, after assuring oneself that the uv artifacts are minimized by choosing 
$\Lambda\geq\Lambda^{NN}$. 

We have shown   earlier  that  extrapolations with the cutoff $\hbar\omega\propto\lambda^2$ work equally well to remove ir artifacts as does an extrapolation with $\lambda_{sc}$.  Consider the example of the nucleus of $^6$He calculated with the JISP16 potential.  The removal of ir artifacts by sending $\hbar\omega\rightarrow 0$  (Figure 10 of Ref.~\refcite{Coon2012}) or with $ \lambda_{sc} \rightarrow 0 $  (Figure 11 of Ref.~\refcite{Coon2012}) takes a different trajectory for each fixed $\Lambda\geq\Lambda^{NN}$   but each extrapolation ends up at the same final value free of model space ir and uv artifacts.  These final values are -28.78(50) MeV from the ir energy  cutoff $\hbar\omega$ and -28.68(22) MeV  from the ir momentum cutoff $\lambda_{sc} \propto \sqrt {\hbar\omega/N}$.  Both results are in agreement with an extrapolated value of -28.70(13) MeV from the extrapolation of a hyperspherical harmonics expansion\cite{EIHH} and with the -28.69(5) MeV or -28.68(12) MeV from the traditional extrapolation of Eq.~\ref{eq:traditional} published in Ref.~\refcite{Maris09}.  A similar demonstration of these very different regulators with the same final result for  $^3$H and Idaho N$^3$LO is given by  Figures 6 and  7 of Ref.~\refcite{Khabarovsk}.  For that case the final result of six  extrapolations with the ir regulator  $\hbar\omega$ is a mean of -7.832 MeV and standard deviation of 0.020 MeV which is in good agreement with the mean of -7.8523 MeV and standard deviation of 0.0008 MeV obtained with the ir regulator $\lambda_{sc}$.  Both results are in good agreement with the consensus values of -7.855 MeV from a 34-channel Faddeev calculation\cite{BNV} and -7.854 MeV from a hyperspherical harmonics expansion\cite{EIHH}.    

 Nevertheless,  Figure 2 demonstrates that the ir cutoff $\lambda_{sc}$ is independent of the uv cutoff $\Lambda\geq\Lambda^{NN}$  and, henceforth, we will take the infrared cutoff to be $\lambda_{sc}$ itself.  As has often been emphasized  in EFT, the choice of regulator is a matter of convenience.  From that perspective the continued refinements of the exact values  of the ir regulator parameter in Refs.~\refcite{later1,later2,later3} seem somewhat beside the point.    

Thus an  extrapolation of the ground state energy with the infrared cutoff $\lambda_{sc}$ is performed by a fit of an exponential plus a constant to each set of results at all  $\Lambda \geq \Lambda^{NN} $.  That is, we fit the ground state energy with three adjustable parameters using the relation 

 \begin{equation}
  E_{gs}(\lambda_{sc}) =  {\cal A} \exp(-{\cal B}/\lambda_{sc}) + E_{gs}(\lambda_{sc}=0).    \label{eq:irextrap}
  \end{equation}     

Incidentally, the traditional choice of $\hbar\omega$ at the bottom of the curves of Figure 1 is plausible  for an extrapolation in $N$ at fixed $\hbar\omega$ but it becomes less attractive when one considers the uv and ir aspects of the traditional extrapolation.
 At fixed  $N$ one does remove the  infrared artifacts by lowering the infrared cutoff ($\lambda_{ir} \propto \sqrt{\hbar\omega}$) but actually increases the uv artifacts because lowering $\hbar\omega$ also lowers the ultraviolet cutoff ($\Lambda \propto \sqrt{\hbar\omega}$).  The loss of uv physics due to the  lower $\hbar\omega$ overwhelms the gain of ir physics and the estimate of the ground state becomes very bad.  A similar situation holds as $\hbar\omega$ increases:  the uv cutoff increases toward $\infty$ so that more uv physics is captured but the ir cutoff also rises and more and more of the infrared artifacts appear.   At the minimum of the $N=N_{max} + 1 = 17$ curve the variational parameters are nowhere near their limits in the ($\Lambda,\lambda_{ir}$) regulator picture and the variational energy is not very good.   Because  $N \propto \Lambda/\lambda_{sc}$, increasing the truncation parameter $N$ simultaneously increases the uv cutoff and decreases the ir cutoff so that the curves move lower  and lower.   Nevertheless,  attempting an extrapolation in $N$ at fixed $\hbar\omega$ chosen at the minimum is less reliable  than the extrapolation techniques examined in this paper  because the extrapolated value depends upon $\hbar\omega$ as seen in Figure 1.

  \begin{figure}[!t]
       \centering
        \includegraphics[width=0.90\textwidth]{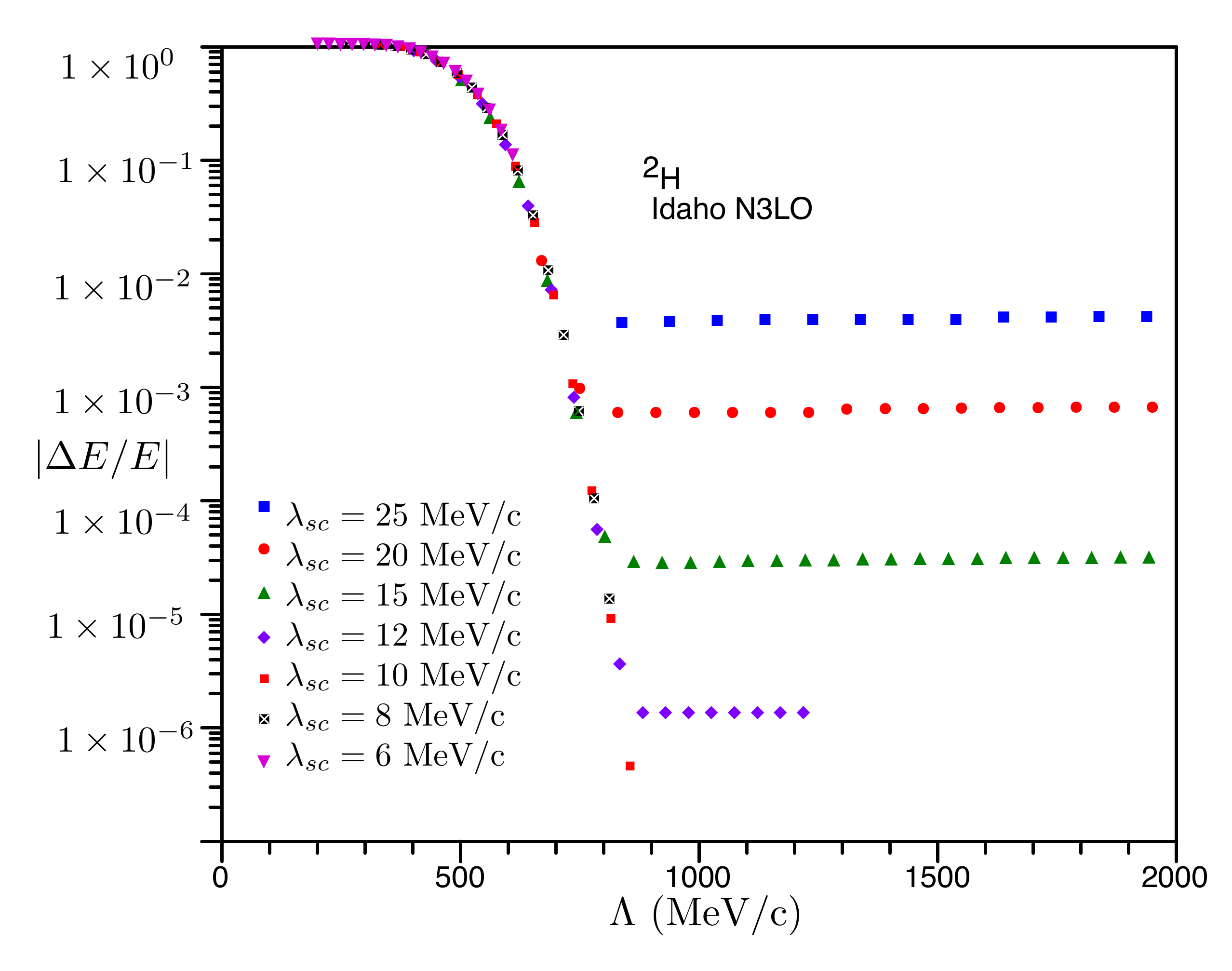}
\caption{(Color online)  Dependence of the ground-state energy of $^2$H upon the uv momentum cutoff   $\Lambda$ for different values of  the ir momentum cutoff $\lambda_{sc}$.   }  
\label{fig:4}
\end{figure}

In Figure 4 we hold fixed the ir cutoff of ($\Lambda,\lambda_{ir} \equiv\lambda_{sc} $) to display the running of  $\vert\Delta E/E\vert$ upon the cutoff $\Lambda$.    Again  a universal pattern at  low $\Lambda$ and plateaus at higher $\Lambda$ are evident.   
For fixed $\lambda_{sc}$, $\vert\Delta E/E\vert$ does not go to zero with increasing $\Lambda$, and indeed even appears to rise for  fixed $\lambda_{sc}\geq 12$ MeV/$c$ and $\Lambda\geq 800$ MeV/$c$. 
Such a plateau-like  behavior was attributed in Figure 2  to a uv regulator scale characteristic of the $NN$ interaction.  (It has been suggested that these plateaus are due to numerical inaccuracies in the HO two-body matrix elements because of the high values of $\hbar\omega$ required for large $\Lambda$.
On the contrary, the values of $\hbar\omega$ needed in Figure 4 are in the range (22-52 MeV) for the topmost curve labeled $\lambda_{sc} = 25$ MeV/$c$ and are lower for the remainder of the curves.  This is the range that NCSM calculations use regularly.)  Another ``missing contributions" argument leads to  a universal behavior at low $\Lambda$  only if $\lambda_{sc}\leq\lambda^{NN}_{sc}$ where  $ \lambda^{NN}_{sc}$ is a second characteristic ir regulator scale implicit in the $NN$ interaction itself.  One can envisage such an ir regulator scale  as related to the lowest energy configuration that the $NN$ potential could be expected to describe.  This would be in the range of the deuteron binding momentum $Q=46$ MeV/$c$  down to about  16 MeV/$c$ which is the (inverse of the) average of the four $NN$ scattering lengths.  However the behavior of the running as  $\Lambda \geq \Lambda^{NN}$   again suggests that corrections are needed to  $\Lambda$ and $\lambda_{sc}$ which are presently defined only to leading order in $\lambda_{sc}/\Lambda$.

   Can one make an  estimate of the uv and ir regulator scales of the $NN$ interactions used in these nuclear structure calculations?  It is easy with the JISP16 potential  \cite{Shirokov07}.  
   The $\it s$-wave parts of JISP16 potential are fit to data in a HO space of $N=8$ and $\hbar\omega = 40$ MeV. 
Nucleon-nucleon interactions are defined in the relative coordinates of the two-body system so one should calculate $\Lambda^{NN} =\sqrt{m (N + 3/2)\hbar\omega} $ with the $\it reduced$ mass $m$ rather than the nucleon mass $m_N$ appropriate for the single-particle states of the model space.
   Taking this factor into account, one finds  $\Lambda^{JISP16}\sim$ 600 MeV/$c$ and $\lambda_{sc}^{JISP} \sim 63$ MeV$c$.   In practice, the  uv region seems already captured at $\Lambda > 500-550$ MeV/$c$ \cite{Coon2012}.   The Idaho N$^3$LO interaction was fit to data with a  high-momentum cutoff of the ``super-Gaussian" regulator set at $\Lambda_{N3LO} = 500$ MeV/$c$ \cite{IdahoN3LO}.  What is the uv regulator scale of the Idaho N$^3$LO interaction  appropriate to the discrete HO basis of this study?  A published emulation of this interaction in a harmonic oscillator basis uses  $\hbar\omega = 30$ MeV and $N_{max} \approx 2n=40$ \cite{EIHH}. 
A more systematic study of emulations  gave a few more sets of   $(N,\hbar\omega)$ which describe the $^3$He  ground state energy equally well  \cite{Barnea2}. These successful emulations of the  Idaho N$^3$LO interaction in a HO basis suggests that $\Lambda^{N3LO}\sim$ 900-1100 MeV/$c$ and $\lambda_{sc}^{N3LO}\sim 21-42$ MeV/$c$,  consistent with  figures 2 and 4.  In practice, from calculations of a variety of light nuclei, the uv region of the Idaho N$^3$LO interaction seems already captured at $\Lambda > 800$ MeV/$c$ \cite{Coon2012}.

The intrinsic uv regulator scales of the Idaho N$^3$LO and JISP16 $NN$  differ by as much as 500 MeV/$c$ and the intrinsic uv regulator scale of the  potential AV18 of the Argonne group \cite{AV18} is so high that we saw no sign of a universal curve for the AV18 deuteron, like that of Figure 2, for values of $\Lambda$ up to 1600 MeV/$c$.  On the other hand, the (less well established) intrinsic ir regulator scales of these two potentials are rather closer to each other  and the analogous figure (not shown)   for the AV18 deuteron does have some similarities to the Idaho N$^3$LO deuteron of Figure 4.
This is not surprising and can be related to the fact that experiments only probe a finite region of momenta. The effective uv character of a potential is determined by how one constructs the potential; that is, it depends on the energy range that it was fit to and what physics assumptions went into its derivation.  The potential makers have no knowledge of the high energy behavior that the potential attempts to describe as they attempt to impose a regulator to suppress the unknown high-momentum behavior.  On the other hand, all $NN$ potentials are expected to describe equally well  the measured low energy behavior, be it zero-energy scattering lengths or the energy of the two-body bound state.  So the ir regulator scales  of different $NN$ potentials would be expected to be similar.

How similar should the ir regulator scales  of different $NN$ potentials be?  The basic idea is that the ir regulator scale  $\lambda_{sc}^{NN}$ should be approximately potential independent and somewhere around 30~MeV/$c$ which is the average of the $^ 3S_1$ and the $^1S_0$ scattering lengths.   It is difficult to pin it down further because each potential chooses to fit slightly different zero-energy observables and even the most important observables in the T=1 channel differ by effective range effects because the deuteron is not at zero energy.  That is, the binding momentum  of the deuteron is  46 MeV/$c$ and the inverse of the concomitant scattering length is also 46~MeV/$c$ if the effective range expansion is truncated at the level of the scattering length.  The difference between 46 MeV/$c$ and the 36 MeV/$c$ from a fit at threshold is an effective range effect which is of natural size (the inverse of the pion scale).  The point is that no matter what you fit, your potential needs to incorporate a small momentum scale, but you cannot pinpoint it exactly because it depends on how far you go incorporating effects of natural size.  This type of uncertainly is not unique to potential building but has long been known in EFT treatments: the convergence of low-energy deuteron observables is markedly improved if one fits  the EFT low energy constants to the deuteron binding energy and asymptotic S-state normalization rather than to the $^ 3S_1$ scattering length and effective range \cite{Rupak}.  But, as already mentioned, these two fit choices are equivalent up to the higher-order effective terms and can be taken into account in an uncertainty quantification of the EFT type \cite {FPW}.

\section{Extrapolations}

\begin{figure}[htpb]
\centerline{\includegraphics[width=1.0\textwidth]{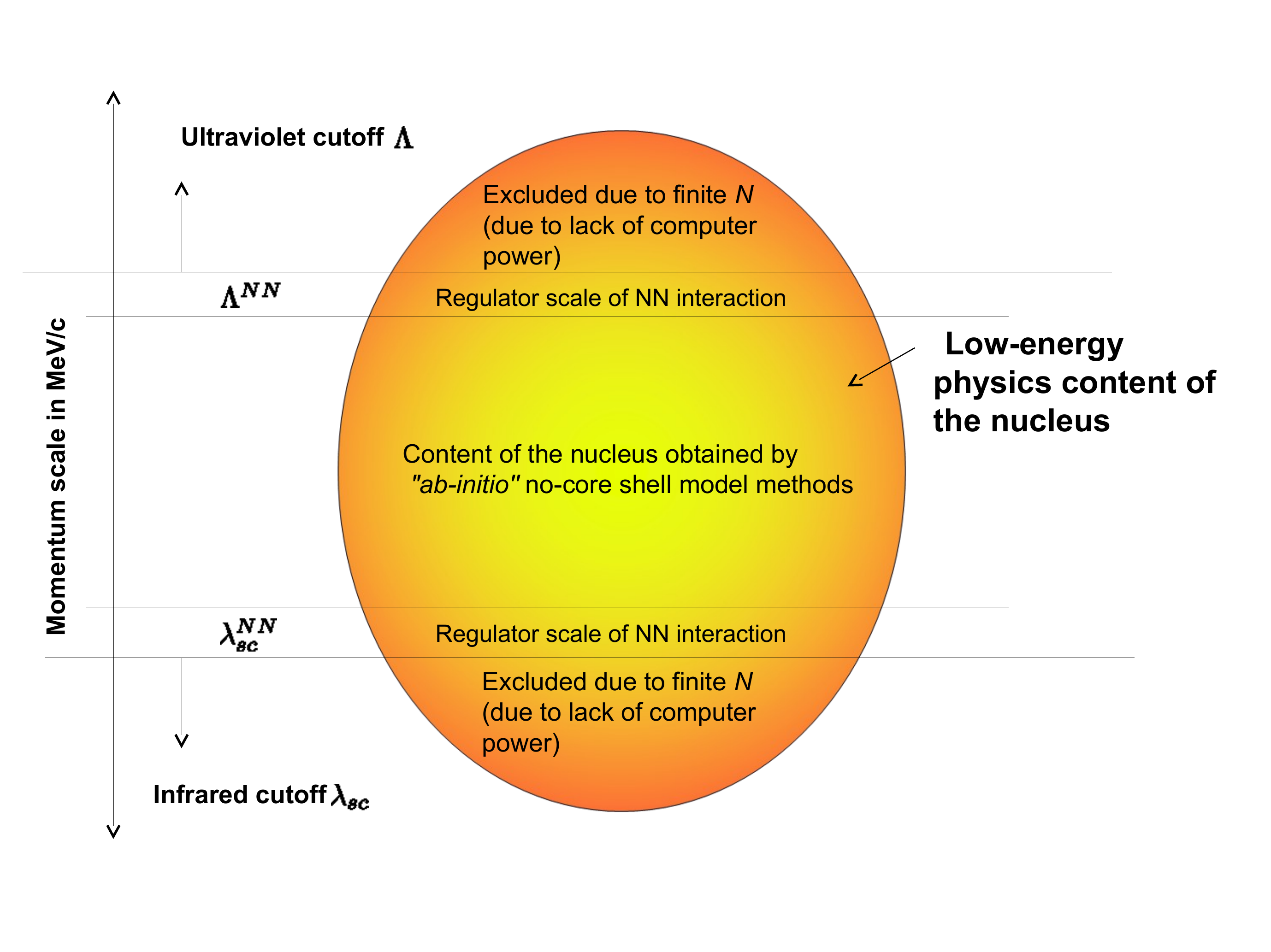}}
\caption{(Color online)  Schematic view of a finite model space (limited by the basis truncation parameter $N$ as described in the text), in which the uv and ir momentum cutoffs are arbitrary.  To reach the full many-body Hilbert space, symbolized by the complete oval,  one expects to let the uv cutoff $\rightarrow \infty$ and the ir cutoff $\rightarrow  0$.  See the text for a discussion of the momentum scales intrinsic to the chosen $NN$ interaction.  }
\label{fig:5}
\end{figure}

\subsection{Regulator scales  and convergence}

The extrapolation scheme proposed in Ref.~\refcite{Coon2012} gives $N$ and $\hbar\omega$ equal roles by employing uv and ir  cutoffs which which should be taken to infinity and to zero, respectively to achieve a converged result (see Figure 5). From figure 2 we conclude uv cutoff $\Lambda=\sqrt{m_N(N+3/2)\hbar\omega}$ should be greater than the intrinsic $\Lambda^{NN}$ of the $NN$ interaction.  Figure 4 suggests that  the ir cutoff $\lambda_{sc}=\sqrt{(m_N\hbar\omega)/(N+3/2)}$ should be less than the intrinsic $\lambda^{NN}_{sc}$ of the chosen $NN$ interaction.  These intrinsic uv and ir scales of the $NN$ interaction are indicated schematically on Figure 5.  Figure 5 allows one to visualize the  extrapolations performed in Ref. \refcite{Coon2012}. There one learned that it was not necessary to take the uv cutoff to infinity.  Instead the uv physics on the top of the oval could be captured by binning all values of the uv cutoff $\Lambda > \Lambda^{NN}$ .With that stipulation,  a single extrapolation in the ir cutoff $\lambda_{sc}$ toward the bottom of the shaded oval caught the ir physics and achieved an extrapolated result which agreed with independent calculations.   The converse extrapolation in the uv cutoff toward the top of the shaded oval with the ir physics expected to be captured by binning or fixing values of $\lambda_{sc} < \lambda^{NN}_{sc}$ was not attempted in Ref. \refcite{Coon2012} but will be addressed later in this paper.


\begin{table}[htpb]
\caption{Intrinsic regulator scales determine  $N$ and $\hbar\omega$ for a converged result}
\label{tab1}  
 \begin{center}
\begin{tabular}{lll}
\multicolumn{3}{ c }{$\Lambda = \Lambda^{NN}\approx 800$ MeV/$c$} \\
\hline\noalign{\smallskip}
$\lambda^{NN}_{sc} \approx 10$ MeV/$c$ & $\lambda^{NN}_{sc} \approx 20$ MeV/$c$ & $\lambda^{NN}_{sc} \approx 40$ MeV/$c$  \\
\noalign{\smallskip}\hline\noalign{\smallskip}
$N \ge 78$ & $N \ge 38$ & $N \ge 18$ \\
$\hbar\omega \le 8$ MeV &$\hbar\omega \le 17$ MeV & $\hbar\omega \le 34$ MeV \\
\noalign{\smallskip}\hline\\
\multicolumn{3}{ c }{$\Lambda = \Lambda^{NN} \approx 500$ MeV/$c$} \\
\hline\noalign{\smallskip}
$\lambda^{NN}_{sc} \approx 10$ MeV/$c$ & $\lambda^{NN}_{sc} \approx 20$ MeV/$c$ & $\lambda^{NN}_{sc} \approx 40$ MeV$c$ \\
\noalign{\smallskip}\hline\noalign{\smallskip}
$N \ge 48$ & $N \ge 23$ & $N \ge 11$ \\
$\hbar\omega \le 5$ MeV &$\hbar\omega \le 11$ MeV & $\hbar\omega \le 21$ MeV \\
\noalign{\smallskip}\hline
\end{tabular}
\end{center}
\end{table}

In this section, however, we keep the uv cutoff $\Lambda > \Lambda^{NN}$ and examine a single extrapolation in the ir cutoff $\lambda_{sc}$.  Noting that $N= \Lambda / \lambda_{sc} - 3/2$ and 
 $\hbar\omega = (\Lambda\lambda_{sc})/m_N$, one can establish the minimum values of $N$ and  the maximum values of $\hbar\omega$ needed for a converged result (see Table I).  The intrinsic $\lambda^{NN}_{sc}$ corresponding to the lowest energy configuration of two nucleons is not well determined by previous numerical investigations of nuclei $A$=2, 3, and 4  (see Figures 4 and 8 of Ref.~\refcite{Coon2012}), so we include a range of values in Table \ref{tab1}. As an example from Table \ref{tab1}, we find that if the intrinsic UV and ir regulator scales are $\Lambda^{NN}\approx 500~$MeV/$c$ and $\lambda_{sc}^{NN} \approx 40~{\rm MeV}/c$, respectively, then one can perform calculations that are converged provided $N \ge 12$ and $\hbar\omega \le 20$~MeV. One can readily adjust  $\hbar\omega$ in a calculation.  It is a computational challenge, however,  to  increase $N$ which gets harder the more particles there are in the nucleus. This is because the many body basis is composed of Slater determinants which are built by distributing $A$ nucleons over a set of single particle states. The number of single particle states steadily increases as the number of oscillator shells increases ({\em i.e.} as $N$ increases). The computational difficulty lies in the fact that the number of Slater determinants grows as a combinatorial factor involving $A$ and the number of single particle states.  This growth is nearly exponential; thus the computational challenge. From Table 1 one concludes that one must extrapolate for all but the lightest nuclei and the softest of interactions.
This fact is related to  the popularity of unitary transformations which do soften original  $NN$ interactions {\cite{BFP,Bognerreview,Jurgenson, FHeblerreview,debate}.

 \subsection{Infrared extrapolations of expectation values of long-range operators}

We now  
extend the extrapolation procedure of Ref.~\refcite{Coon2012} to expectation values of  operators needed to calculate other properties of nuclei.  Such operators include the root mean square ({\it rms})  point radii related, in a model-dependent way,  to the measured size of nuclei and the long-range dipole operator $D$ which governs the electric dipole polarizabilities and total photoabsorption cross section of light nuclei.  Point nucleon radii of nuclei are calculated with the operator  $r^2$ and do not take into account the electromagnetic size of the constituents.  However,  unlike the Hamiltonian, $r^2$  is not  a bounded operator  \cite{Polyzou} and therefore has no convergence theorems with 
$\mathcal{N}$ 
 \cite{Delves72}.  The running of  the expectation value of  $r^2$ with $\hbar\omega$ at fixed $N$ has been contrasted  with  the running with $N$ at fixed $\hbar\omega$,  with hard to interpret convergence results for $A\geq 6$ nuclei \cite{Cockrell12,Maris13}.   An extensive NCSM study in the helium isotopes, calculated with the JISP16 potential, of the runnings of this long-distance operator with respect to $\hbar\omega$  and/or $N$ has now been made \cite{Caprio_cs}.  Calculations were performed  with both the harmonic oscillator basis and a basis of Coulomb-Sturmian radial functions which have the exponential asymptotics expected to be more suitable for nucleons bound by a finite-range force.  
 Early NCSM studies involving the (also unbounded) dipole operator $D$ utilized the phenomenological potentials inspired by chiral perturbation theory and the traditional extrapolation which lets $N\rightarrow\infty$ at a chosen fixed  $\hbar\omega$ \cite{Stetcu1, SofiaPetr}.  A convergence analysis of the traditional extrapolation of Eq.~\ref{eq:traditional}  for electric dipole polarizabilities of $^3$H, $^3$He and $^4$He showed faster convergence for a fixed $\hbar\omega$ lower than that used for  the binding energy itself \cite{Stetcu2}.  
 
  Here we apply, to these long-range operators,  the infrared-ultraviolet  extrapolation procedure of  Ref.~\refcite{Coon2012} which employs  $N$ and $\hbar\omega$ on an equal footing.  We begin with  the definitive point proton radius of   1.436 fm  for $^4$He from the fully  converged calculations of Ref.~\refcite{Caprio_cs}.   In figure 6 we display the running of the {\it rms} point radius  with respect to  $\lambda_{sc}$ for values of $\Lambda$ such that the uv artifacts are minimized by choosing 
$\Lambda\geq\Lambda^{NN}$. We observe that the radius  is exponential in the inverse of  $\lambda_{sc}$ like the running of the ground state energy in Figure 2.  Thus one can use an extrapolation scheme similar to that of Equation~\ref{eq:irextrap}.   This result is quite different from the  radius extrapolation formulae suggested in Ref.~\refcite{FHP} and later derived for one particle in three dimensions in Ref.~\refcite{later2}.   These formulae display  a two term polynomial \cite{FHP} or a three-term polynomial\cite{later2} in odd powers of     ${\cal B}/\lambda_{sc}$ multiplying the exponential in  $-{\cal B}/\lambda_{sc}$ with the admonition that for a robust extrapolation one needs ${\cal B}/\lambda_{sc}$ large enough so that the cubic correction dominates the subleading terms.  From Figure 6 one can see no evidence in this many-body system for such a cubic or linear correction to a simple exponential.

\begin{figure}[htpb]
\centerline{\includegraphics[width=0.90\textwidth]{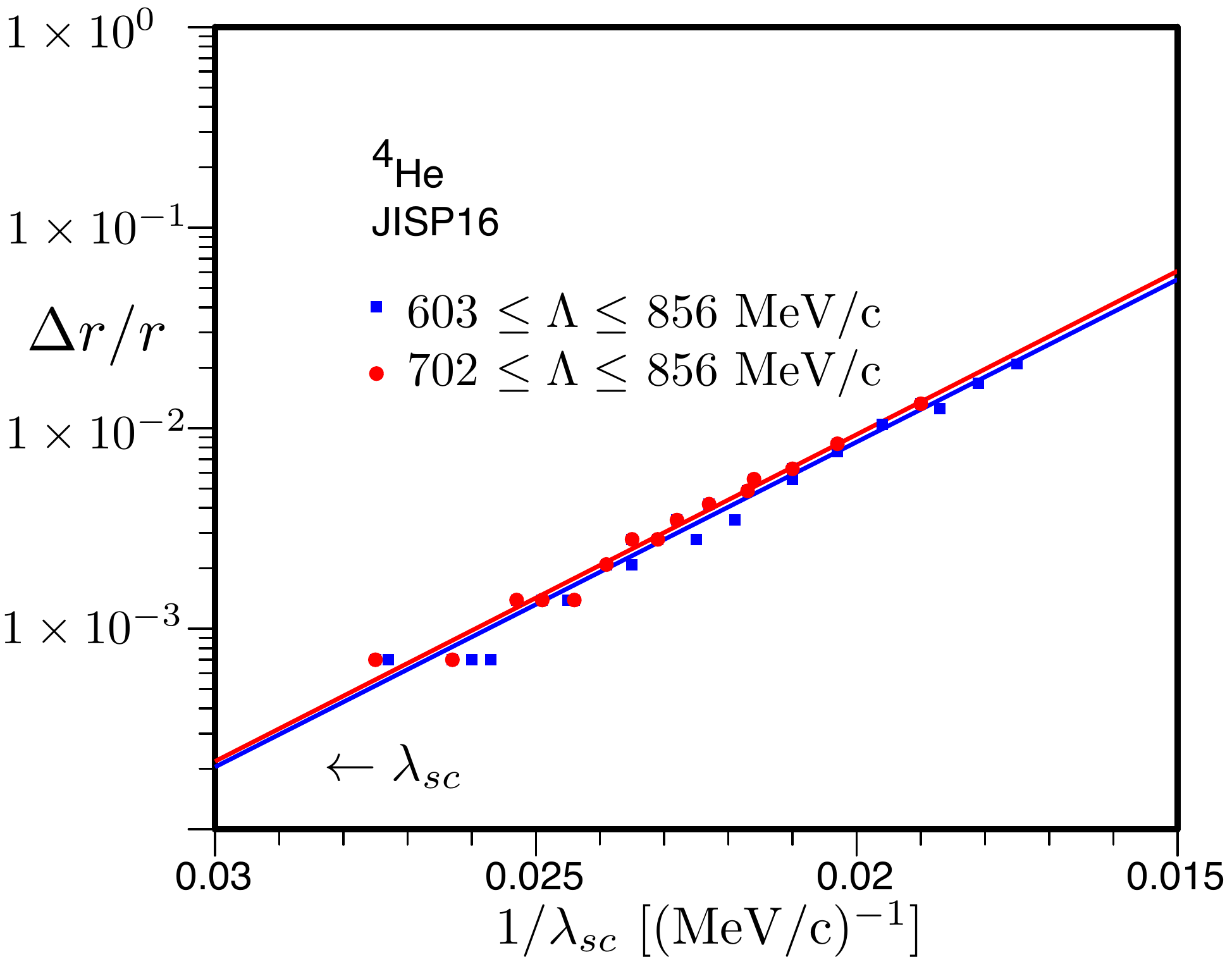}}
\caption{(Color online)    Dependence of the  {\it rms} point radius $\langle 0|r^2|0\rangle^{1/2}$ of $^4$He  upon  the ir momentum cutoff  $\lambda_{sc}$ for $\Lambda$ above the $\Lambda^{NN}$ set by the $NN$ potential.  Here the $NN$ interaction is the JISP16 of Ref. 15.  The curves are exponential fits to the calculated points. $\lsc$ decreases in magnitude from right to left as indicated by the arrow. } 
\label{fig:6}
\end{figure}

In Figures 7, 8  and 9,  we  plot the ground state energy eigenvalue, the {\it rms} point radius, and the total dipole strength of $^4$He  obtained by a NCSM calculation \cite{Micah}, done in a translationally invariant basis which depends only on Jacobi coordinates \cite{NKB}. The $NN$ interaction is the Idaho N$^3$LO  softened by the similarity renormalization group (SRG) evolution according to the method described in Ref.~\refcite{Jurgenson}.
Transforming the Hamiltonian induces  higher order many-body forces which should be kept to preserve the unitary nature of the transformation.  If they are not kept, results become dependent on the SRG flow parameter  which is commonly taken as a momentum parameter that  starts at infinity and approaches zero as the low and high momentum sectors of the interaction partially decouple.  

 It is of interest to learn if the scaling behavior apparent of ground state energies in Figure 2 and the many examples in Ref.~\refcite{Coon2012} is also true for the induced many-body forces  and the  three-body forces added to the Hamiltonian (see Refs.~\citen{Jurgenson, Micah} for a full description of the SRG scheme and nomenclature).  The initial added $NNN$ force was also inspired by chiral perturbation theory  and takes a N$^2$LO form \cite{Bochum} with the two-pion-exchange terms taken from pion-nucleon scattering data and the strengths of shorter range terms fitted to properties of $A=3$ nuclei \cite{Gazit}. 
 For this exercise, we utilized calculations \cite{Micah} with $\hbar\omega$ = 22 and 28 MeV and $N \leq 18$. The SRG momentum parameter $\lambda_{SRG}$ was 1.8 fm$^{-1}$ and our own study of their results suggest that the intrinsic uv cutoff of this SRG transformed $NN$ interaction is  $\Lambda^{NN} \leq 440$ MeV; we therefore group together all $\Lambda \geq\Lambda^{NN}$ to guarantee  capture of the uv physics.    Within this model space      $8\leq N \leq18$ guarantees  $\Lambda \geq\Lambda^{NN}$ for these values of $\hbar\omega$ but, according to  Table 1, values of  $\hbar\omega > 21$ MeV require an extrapolation  in $\lambda_{sc}$. The calculations used the bare $r^2$ and dipole operators, rather than operators transformed using the same unitary transformation as the Hamiltonian.  A later study employs evolved (scalar) operators which do display the unitary nature of the SRG evolution \cite{MicahOperator}, but we are addressing convergence issues here so this advancement in {\itshape ab initio}  technology  should not alter our conclusions.

\begin{figure}[htpb]
\centerline{\includegraphics[width=0.90\textwidth]{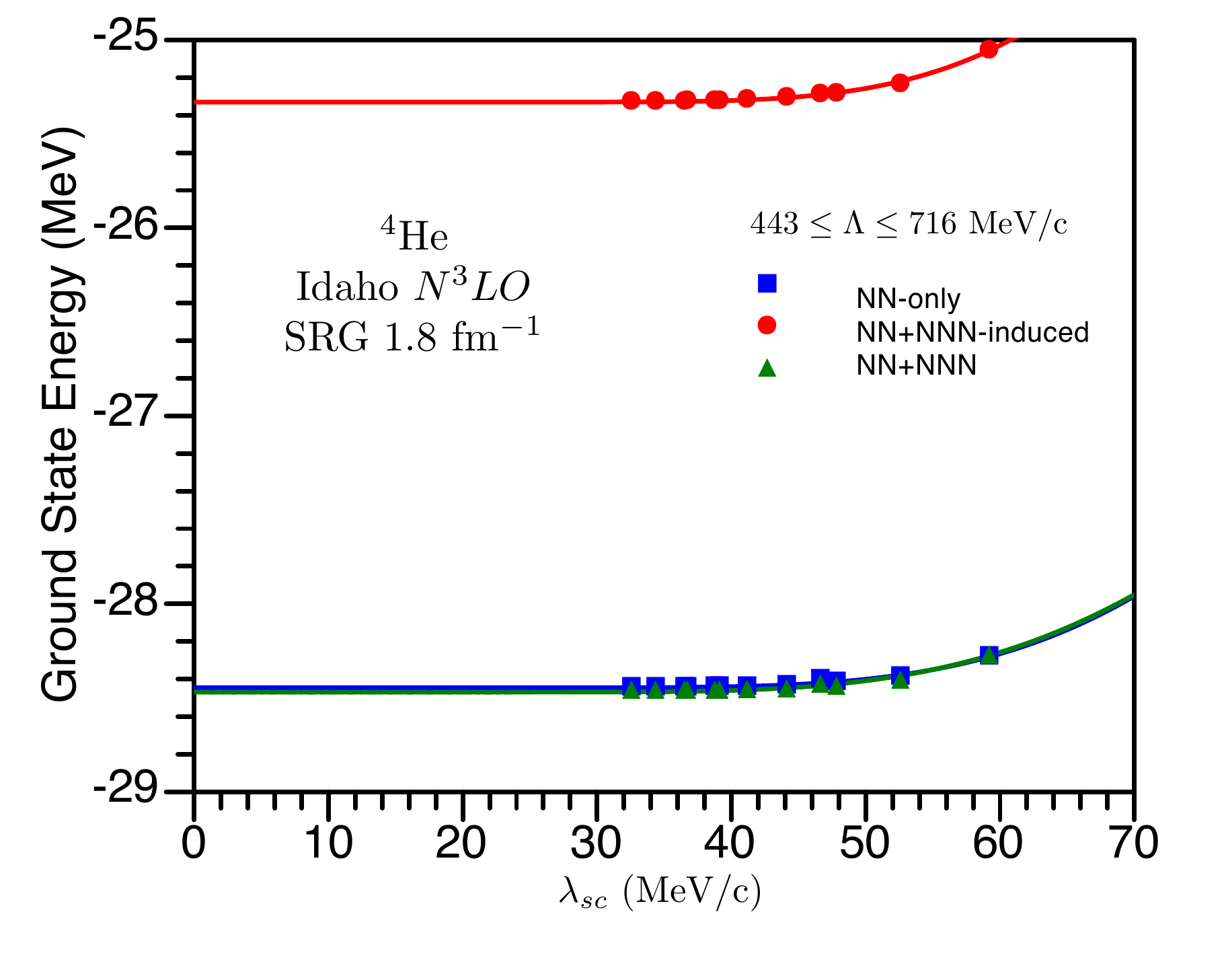}}
\caption{(Color online)  The ground state energy of  $^4$He calculated at values  of $\Lambda\geq 443$ MeV/$c$  and variable $\lambda_{sc}$. The curves are a fit to Eq.~\ref{eq:irextrap}  and the function fitted is  used to extrapolate to the ir limit $\lambda_{sc}= 0$.  The three SRG transformed Hamiltonians are described in the text.}
\label{fig:7}
\end{figure}

 \begin{figure}[htpb]
\centerline{\includegraphics[width=0.90\textwidth]{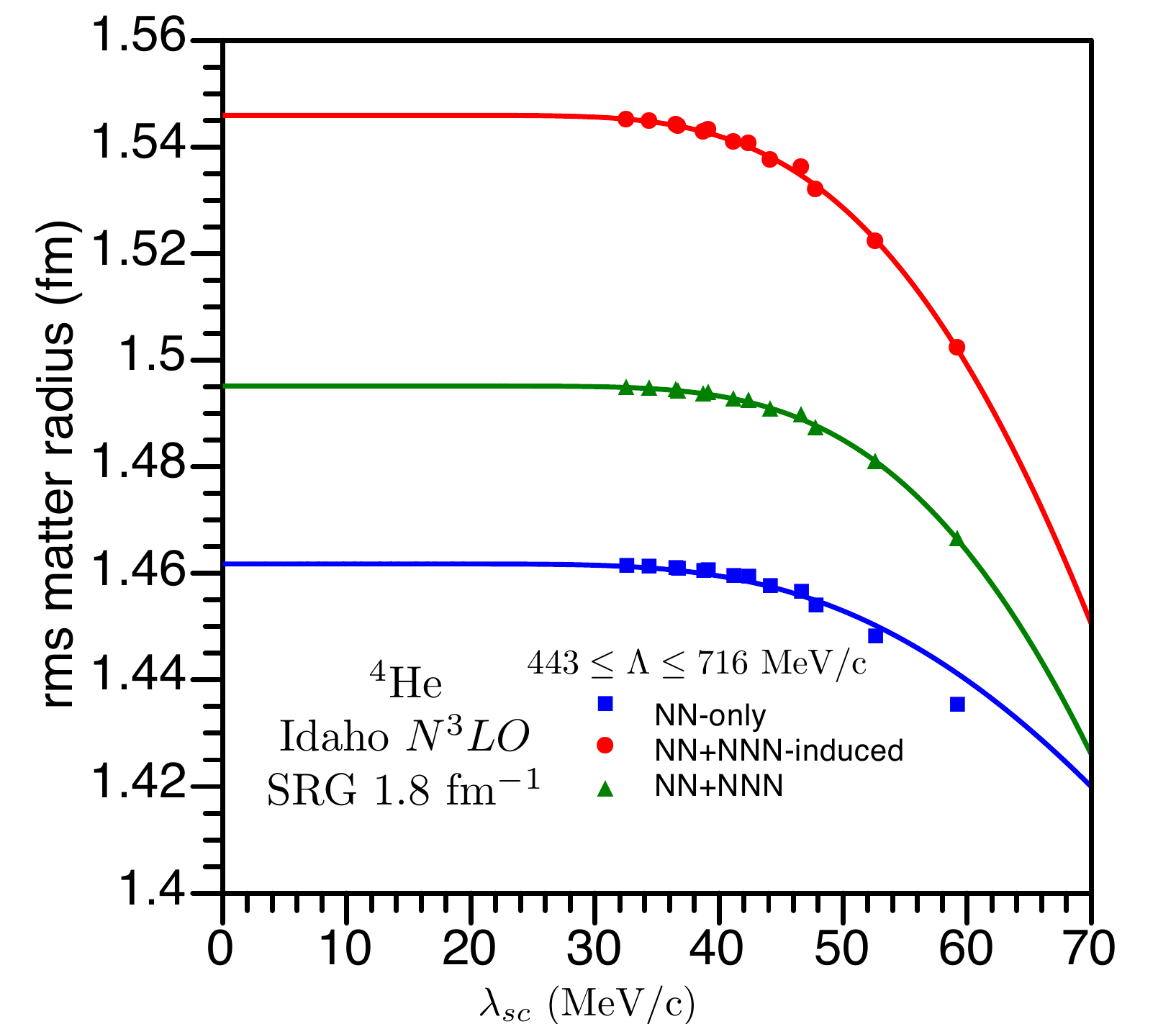}}
\caption{(Color online)  The  {\it rms} point radius $\langle 0|r^2|0\rangle^{1/2}$ of $^4$He calculated as in Figure 7.}
\label{fig:8}
\end{figure}

  \begin{figure}[htpb]
\centerline{\includegraphics[width=1.0\textwidth]{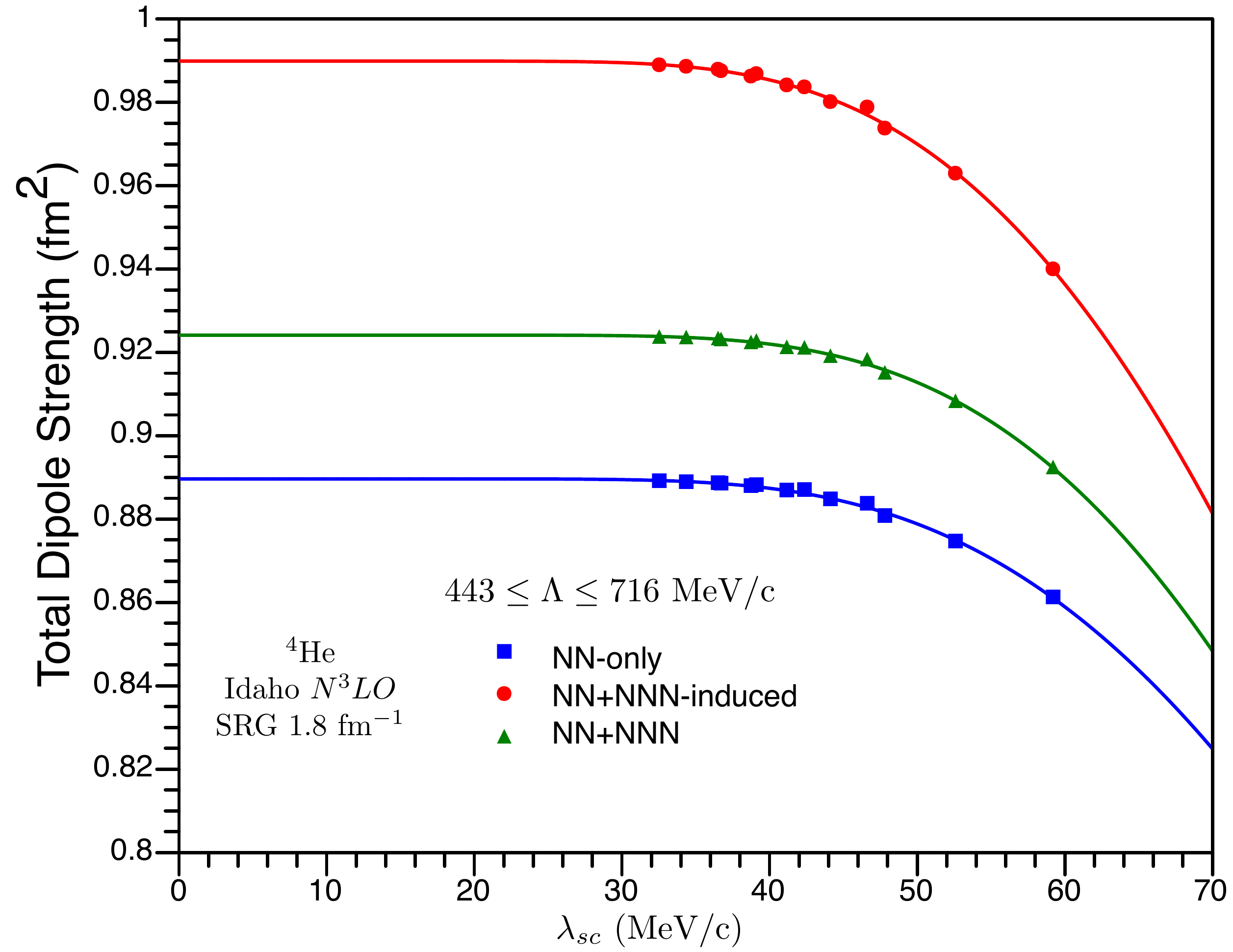}}
\caption{(Color online)  The total dipole strength $\langle 0|\mathbf{D}\cdot\mathbf{D}|  0\rangle$ of $^4$He calculated as in Figure 7. Here  $\mathbf{D}$ is the unretarded dipole operator  defined in Ref.~\cite{Micah}.   }
\label{fig:9}
\end{figure}

The extrapolation is performed by a fit of an exponential plus a constant to each set of results at all  $\Lambda \geq \Lambda^{NN} $.  That is, we fit the ground state energy with three adjustable parameters using   Eq.~\ref{eq:irextrap}. 
 The {\it rms} point radius and the total dipole strength are obtained by similar fits: 
 \( r(\lambda_{sc}) =  {\cal A} \exp(-{\cal B}/\lambda_{sc}) + r(\lambda_{sc}=0)  \)
  and
\( D^2(\lambda_{sc}) =  {\cal A} \exp(-{\cal B}/\lambda_{sc}) +D^2(\lambda_{sc}=0)  \).  
In all these formulae, the values of both  ${\cal A}$ and ${\cal B}$ are specific to the expectation value of the operator being calculated. The extrapolation formulae work equally well for the induced three-body forces and the added three-body forces.   The running of the {\it rms} point radius and the total dipole strength with $\lambda_{sc}$  is about the same, because of the long known approximate relationship between them \cite{Foldy}  which is satisfied very well for $^4$He.

It should be noted that our  extrapolations in these figures employ an exponential function whose argument
$1/\lambda_{sc}  =  \sqrt{(N + 3/2)/(m_N \hbar\omega)}$ is proportional to $\sqrt{N/(\hbar\omega)}$.
This  procedure of extrapolating $\lambda_{sc}$ downward from the values allowed by computational limitations treats both $N$ and $\hbar\omega$ on an equal basis.  The exponential extrapolation in $\sqrt{N/(\hbar\omega)}$ is therefore distinct from the traditional  extrapolation which employs an exponential in $N_{max}$ ($=N$ for this $s$-shell case) \cite {NQSB, NavCau04,  Maris09, CP87, Forssen2, Jurgenson, Cockrell12, Maris13, BNV}.   The convergence of all three operators is the same with the $\lambda_{sc}$ extrapolation, in contrast to the traditional extrapolation for the same data ($N\rightarrow\infty$ at fixed  $\hbar\omega$)  which found slower and slower convergence for the ground state energy eigenvalue, the  {\it rms} point  radius, and the total dipole strength \cite{Micah}.  As the model space is large and the intrinsic uv cutoff is small, the extrapolated results obtained here agree with those of the traditional extrapolation used in  Ref.~\refcite{Micah}.

\subsection{Dependence of ir extrapolation upon nucleus and $NN$ interaction}

 \begin{figure}[htpb]
\centerline{\includegraphics[width=0.90\textwidth]{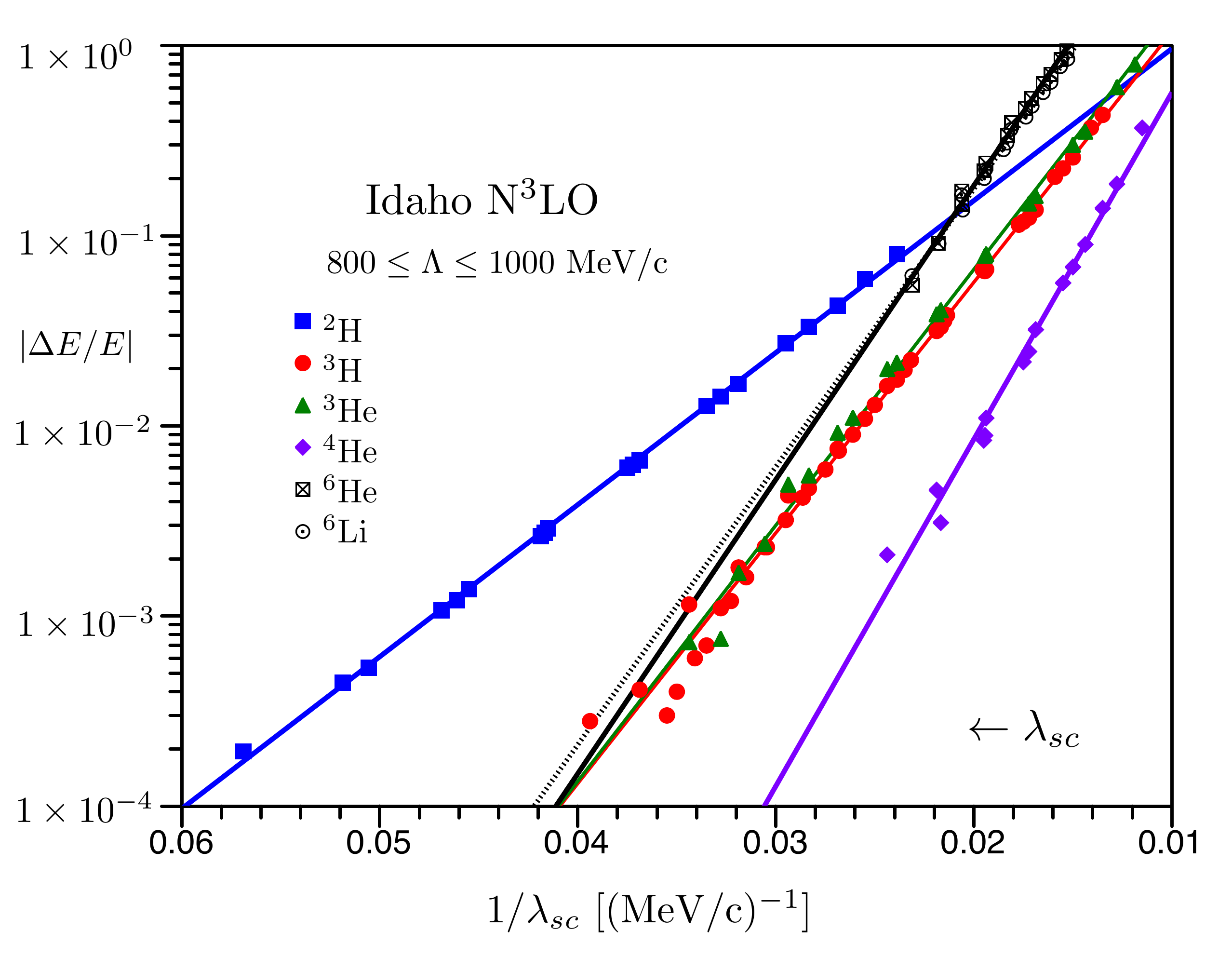}}
\caption{(Color online)  Dependence of the ground-state energy of four $s$-shell nuclei and two $p$-shell nuclei (compared to a consensus value-see text)  upon  the ir momentum cutoff  $\lambda_{sc}$ for $\Lambda$ above the $\Lambda^{NN}$ set by the $NN$ potential. Here the $NN$ interaction is the Idaho N$^3$LO  of Ref.~14. The curves are exponential fits to the calculated points. $\lsc$ decreases in magnitude from right to left as indicated by the arrow.} 
\label{fig:10}
\end{figure}

   \begin{figure}[htpb]
\centerline{\includegraphics[width=0.90\textwidth]{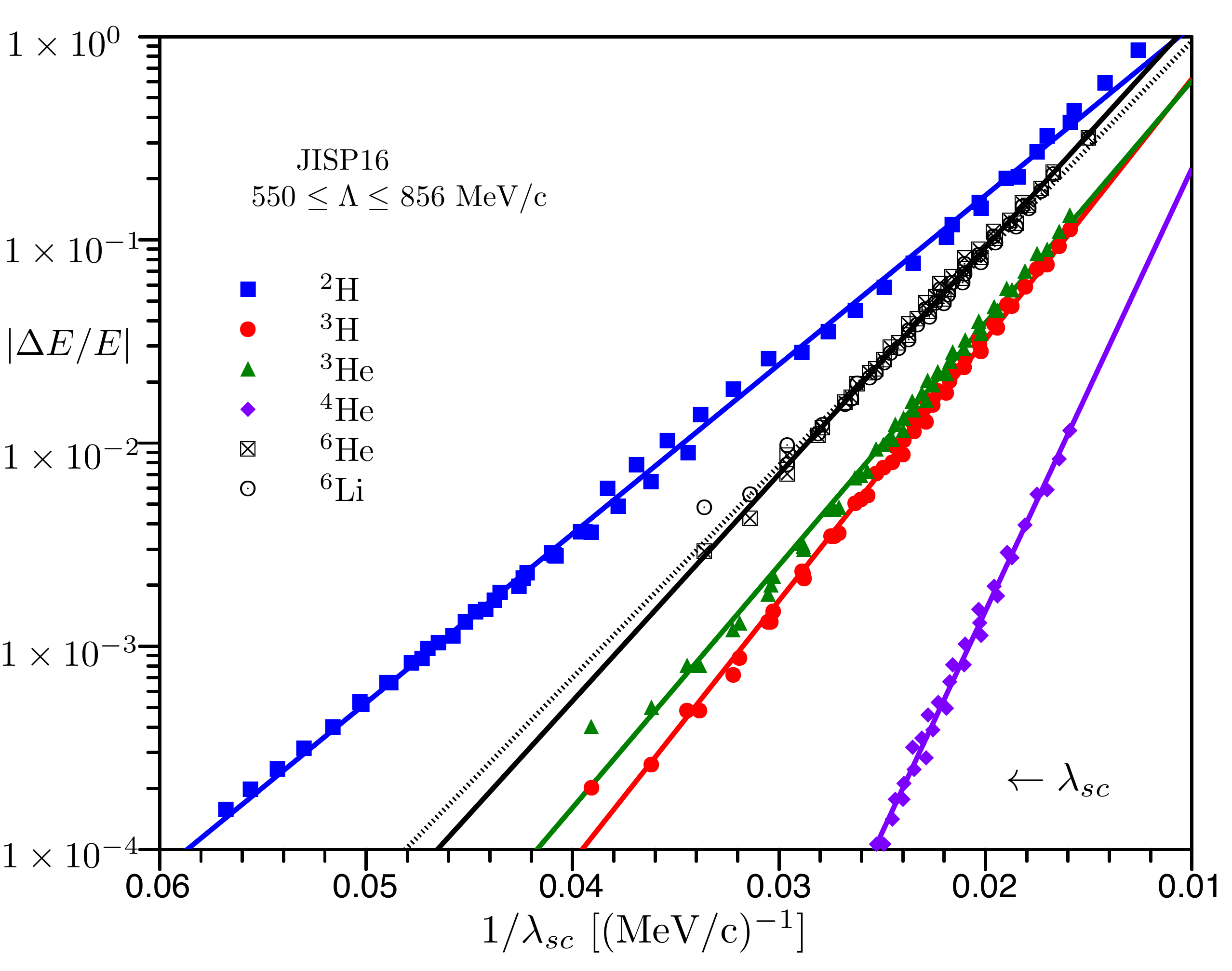}}
	\caption{(Color online)  Dependence of the ground-state energy of four $s$-shell nuclei and two $p$-shell nuclei (compared to a consensus value-see text)  upon  the ir momentum cutoff  $\lambda_{sc}$ for $\Lambda$ above the $\Lambda^{NN}$ set by the $NN$ potential.  Here the $NN$ interaction is the JISP16 of Ref.~15.  The curves are exponential fits to the calculated points. $\lsc$ decreases in magnitude from right to left as indicated by the arrow.}
\label{fig:11}
\end{figure}
 
For Figure  10, we return to calculations with the original Idaho N$^3$LO $NN$ interaction and display a generalization of Figure 2 which includes more massive nuclei in addition to the deuteron.  We  take advantage of the ``capture"  of the uv region by binning all results with $\Lambda\geq 800$ MeV/$c$.  They do indeed fall on a universal curve for each nucleus shown, indicating that one can use this universal behavior for an extrapolation which is somewhat independent of $A$ for $A>2$. The points can be  fit by the function $y = {\cal A}\exp(-{\cal B}/\lambda_{sc})$ with ${\cal B}\approx 200-400$ MeV/$c$ for the s-shell nuclei (see Table 3 for the values of $\cal B$ for each nucleus).  The increase in the value of  $\cal B$ indicates a higher rate of convergence in the ir momentum cutoff as the number of particles is increased up to four.   However, the higher rate of convergence is offset  by the smaller  values of $\lambda_{sc}$ needed for a satisfactory calculation.   The lowest value of $\lambda_{sc}$ available to our calculations is set by  $\lambda_{sc} = \Lambda/(N+3/2)$     where $\Lambda=800$ MeV/$c$, the lowest value which seems to capture the uv physics.  If one
draws an imaginary horizontal line on Figure 10, say at the 1\% level in $\Delta E/E$, one sees that  the loosely bound deuteron (charge {\it rms} radius  $r_{ch} =2.130 \pm 0.010$ fm) requires a smaller value of $\lambda_{sc}$ to capture the ir physics at this level in $\Delta E/E$
than needed for  the more tightly bound triton ($r_{ch} = 1.755 \pm0.087$ fm) and even more tightly bound $\alpha$ particle ($r_{ch} = 1.680 \pm 0.005$ fm) \cite{Sick08}.   The smallest values available correspond to  the farthest left hand points of the figure,  which are  $\lambda_{sc} \sim 25$ MeV/$c$ for the triton calculation (largest $N=30$) and $\lambda_{sc}\sim$ 41 MeV/$c$ for the $\alpha$ calculation (largest $N=18$).   These values of the ir cutoff  can be lowered (thereby increasing the reliability of the extrapolation) only by increasing $N$; a computational challenge which gets harder the larger the number of particles in the nucleus. For example, the largest $N$ achievable with ANTOINE \cite{ANTOINE}  for the  nuclei $^6$Li and $^6$He is 17 ($N=N_{max} +1$ for these p-shell nuclei).  As the value of $\Lambda$ must be 800 MeV/$c$ or greater for  the Idaho N$^3$LO  $NN$ interaction  the smallest value of $\lambda_{sc}$ is then  $\sim 43$ MeV/$c$.  For the softer JISP16 $NN$ potential which has a lower minimum $\Lambda^{JISP16} \approx 500$ MeV/$c$  the ir extrapolation is easier, as demonstrated by Figure 11 in which the calculated points extend much further to the left (i.e., are much lower in $\lambda_{sc}$) for large A.

Notice that the slopes of the light $p$-shell nuclei $^6$He and $^6$Li are intermediate between those of the more massive $s$-shell nuclei and the slope of the deuteron which has an unnaturally small binding energy  on a hadronic scale.
It is tempting to try to understand the slopes in Figures 10 and 11  with the aid of the scale of  $Q$; the experimental binding momentum of each nucleus.  In a non-relativistic EFT it is the binding momentum $Q$ that determines whether a bound state is within the region of validity of the expansion  \cite{urBira}.  The definition of the binding momentum of a two-body bound state is straightforward \cite{X(3872)}  and is often used in  LQCD and EFT  studies which need to take into account the unnaturally small size of the deuteron binding energy \cite{Raul, Beane}.   Unfortunately, the extension of the definition of $Q$ to more massive nuclei is not straightforward.  The analyses in the literature, of which we are aware, provide two alternate forms \cite{Liebig11}; $\bar{Q} = \sqrt{2 m_N(E/A)}$ where $E/A$ is the binding energy per nucleon, or $Q = \sqrt{2 \mu \epsilon}$  where $\mu$ is the reduced mass of a single nucleon with respect to the rest of the nucleons in the nucleus and $\epsilon$ is the binding energy with respect to the first breakup channel.  Clearly the two definitions coincide for the deuteron.  Both give similar values of the binding momentum scale for the $s$-shell nuclei, see Table 2,  but differ significantly for the light $p$-shell nuclei.  With the definition $Q$, the binding momenta of $^6$He and $^6$Li, respectively, are  comparable to that of the deuteron because the first breakup channel into $^4$He$ +2n$ and $^4$He$ +d$, respectively, is only about 1 MeV above the ground state \cite{Tilley,Dataeval}.  In contrast, according to  definition $\bar{Q}$, the binding momenta of these $A=6$ nuclei, which have a tightly bound $^4$He core, are closer to the binding momentum of $^4$He itself. 

 The binding momentum definition $Q$ is attractive because it seems to  reflect the structure of states with valence nucleons nearly decoupled from a core (or cluster) of more tightly bound nucleons.  Such states whose extent is larger than the range of the force are  called halo nuclei.  The obvious example is  the deuteron which may have no core but has both a valence neutron and a valence proton.    $^6$He fits this characterization because the overlap of the NCSM wave function with the translationally invariant three-body channel $^4$He$ +2n$  displays vividly the  cluster structure of  $^6$He \cite{Forssen6He}.  The less weakly bound  $^6$Li has a valence proton and a valence neutron and the overlap of three-body wave functions obtained via Faddeev methods with the $^4$He$ +d$ cluster is about 60-70\% \cite{Coon6Li} implying that the valence proton and valence neutron are uncorrelated as much as 30-40\% of the time in  $^6$Li.

\begin{table}[htpb]
\caption{Nuclear binding momenta and  slopes  (${\cal B}$) (rescaled by the nuclear binding momenta) of $\Delta E/E$ vs $1/\lambda_{sc}$ plots.  Experimental data are from Refs.~\citen{Tilley,Dataeval} }
\label{tabQQbar}  
 \begin{center}
\begin{tabular}{lcccccc}

    nucleus    &   exp.      &      exp.    &    JISP16  & N$^3$LO  &    JISP16  & N$^3$LO\\
       \noalign{\smallskip}\hline
& $Q$ & $\bar{Q}$  &  ${\cal B}/Q$ & ${\cal B}/Q$ &  ${\cal B}/\bar{Q}$ &  ${\cal B}/\bar{Q}$\\
                & (MeV/$c$)   & (MeV/$c$)   &  &    &     &    \\
\noalign{\smallskip}\hline
$^2$H & 46 & 46 & 4.17 & 4.00 & 4.17 & 4.00\\
$^3$H & 88& 73 & 3.36 & 3.45  & 4.05 & 4.34\\
$^3$He & 83 & 70 & 3.31 & 3.73  & 3.92 & 4.42\\
$^4$He & 167 &115  & 3.00 & 2.51 & 4.36 & 3.65 \\
$^6$He &   39  &   96  &6.59 & 9.13 & 2.68 & 3.71 \\
$^6$Li  & 48 & 101 & 5.00  & 7.00 & 2.38 & 3.33 \\
\noalign{\smallskip}\hline

\end{tabular}
\end{center}
\end{table}

 In Table 2, we scale each slope by the value of the putative binding momentum of its nucleus, to learn if  the rescaled slopes would come to a narrower range suggesting a common universal slope or common rate of convergence. The rescaled slopes in the top line corresponding to the deuteron are nearly the same with respect to the alternate  definitions of the binding momentum (identical for the two-body bound state) and with respect to the $NN$ potential, each of which is fit to the experimental binding energy.  The deuteron line suggests that there is little difference in the running of the ground state energy of the deuteron with the two $NN$ potentials, both of which are fit to this on-shell datum. The difference in the right hand columns shows up in the rest of the $s$-shell nuclei.   The rescaled slopes are rather similar for a given potential.  The pattern of the columns can perhaps be understood by the observation that, although both potentials are fit to the deuteron binding energy,
the JISP16 potential has been tuned off-shell \cite{Shirokov07} to  provide good descriptions of $^3$H binding, the low-lying spectra of $^6$Li and the binding energy of $^{16}$O.  The Idaho N$^3$LO potential, however, underbinds $s$-shell nuclei and the light $p$-shell nuclei  \cite{NavCau04, NQSB, BNV} and is often supplemented  by a $NNN$ potential if one wants to describe nature.  Table 2 suggests that,  if replotted, the points of the $A=2,3,4$ nuclei would lie in a band,  the bands differing according to the $NN$ potential.   Not so for the nuclei  $^6$He and $^6$Li.  If one rescales by  $\bar{Q}$  (based upon binding energy per particle) the rescaled slopes of the $A=6$ nuclei remain within the ${\cal B}\approx 200-400$ MeV/$c$ range of the data displayed in Figures 9 and 10.  But if one rescales by $Q$ (based upon binding energy with respect to  the lowest breakup channel) the rescaled slopes of the $A=6$ nuclei are well above that of $^4$He and are real outliers.  We have no speculation for this behavior, but wonder if it suggests a guide for a more definitive definition of nuclear binding momentum.  This definition certainly needs a better theoretical grounding.

\begin{table}[htpb]
  \caption{Single nucleon separation energies and slopes of $\Delta E/E$ vs $1/\lambda_{sc}$ plots.  Separation energies $S$ are in MeV and calculated slopes $\cal B$ are in MeV/$c$.  Experimental data are from Refs.~\citen{Tilley,Dataeval}}    
\label{tab2}  
 \begin{center}
\begin{tabular}{ll|lcc|lcc}

 \multicolumn{2}{l}{nucleus  } & \multicolumn{3}{c} {JISP16} & \multicolumn{3}{c} {N$^3$LO}\\
 \noalign{\smallskip}\hline
  & $S$ exp. & $S$ & $4\sqrt{m_N c^2 S}$  &  $\cal B$  &$S$ & $4\sqrt{m_N c^2 S}$ & $\cal B$ \\
\noalign{\smallskip}\hline
$^2$H & 2.22457 & 2.2246 & 183 & 192 &2.2246 &183 &184\\
$^3$H & 6.25723 & 6.144 & 304 & 296& 5.625 &291 &304\\
$^3$He & 5.49348 & 5.441&286 &275 & 4.890  &271 & 310\\
$^4$He &20.20  &20.28  &552 &502 &17.89  &518 & 420\\
\noalign{\smallskip}\hline

\end{tabular}
\end{center}
\end{table}

There is another way to analyze the running of  the variational ground state energies with the cutoff $\lambda_{sc}$ displayed in Figures 10 and 11 and quantified in Table 3.  As discussed in Section {2, the HO basis of the model space confines the system, in effect, within a finite coordinate space volume.  This feature  has been utilized in later papers \cite{FHP,later1,later2,later3,later4} to    suggest   an ir cutoff in coordinate space larger than the range of the potential.  This cutoff   {\it increases} to remove ir artifacts.   This model cutoff ``{\it L}" relies on a length equivalent to the radius of the hard wall of a spherical confining enclosure.  The  determination of the actual value of ``{\it L}" has been refined in the latter papers \cite{later1,later2, later3,later4} of this series but the hard wall (Dirichlet boundary condition) interpretation  of the ir regulator remains. The numerical value of  ``{\it L}" is found to be quite similar to $\hbar/\lambda_{sc}$ (modulo defining factors of $\sqrt{2}$ and smaller {refinements documented in the later papers).   Extrapolation with this coordinate space cutoff  has been studied extensively in the two-body bound state \cite{later1,later2}.  The derivations there apply to the leading-order ir extrapolation formula for an $s$-wave bound-state energy of a single-particle system.  The conclusions of these studies are: 
``The derivations ...   imply that the energy corrections should have the same exponential form   and  functional dependence on the radius  {\it L} at which the wave function is zero, independent of the potential ... "\cite{later1}.  In Table 3 we compare for the $s$-shell  nuclei our calculations of the running with the ir cutoff   $\lambda_{sc}$ with these expectations of this  coordinate space cutoff procedure.    The analogue of the slope ${\cal B}$ of our formula is a momentum which corresponds to the separation energy $S$ of the least bound valence nucleon (not to be confused with the   binding momentum of the nucleus itself).   Given this analogue it is straightforward to transmogrify the ir extrapolation formula of Ref.~\refcite{FHP} into the form of our ir extrapolation formula Eq.~\ref{eq:irextrap}.} As in Table 2, the running with the ground state energy of the deuteron is the same for both types of ir cutoffs and both $NN$ potentials.    Going on to heavier nuclei we  see that the running does depend  upon the $NN$ potential and that for $A=2$ and $A=3$ nuclei the slopes of Figures 10 and 11 do follow approximately  the separation energies $S$ calculated for each potential.  That is, the  approximation of a three-body wave function by a two-body wave function works for this case, as suggested by the successful baryon-dibaryon formulation of the triton in pionless EFT \cite{PauloBira}.  However, for $A=4$ the significant difference  between the slope expected from the separation energy and the actual slope of the Figures and of Table 3 suggests that the four-particle bound state is not well approximated by a two-body bound state.

\subsection{Ultraviolet extrapolation}

  \begin{figure}[ht]
\centerline{\includegraphics[width=0.90\textwidth]{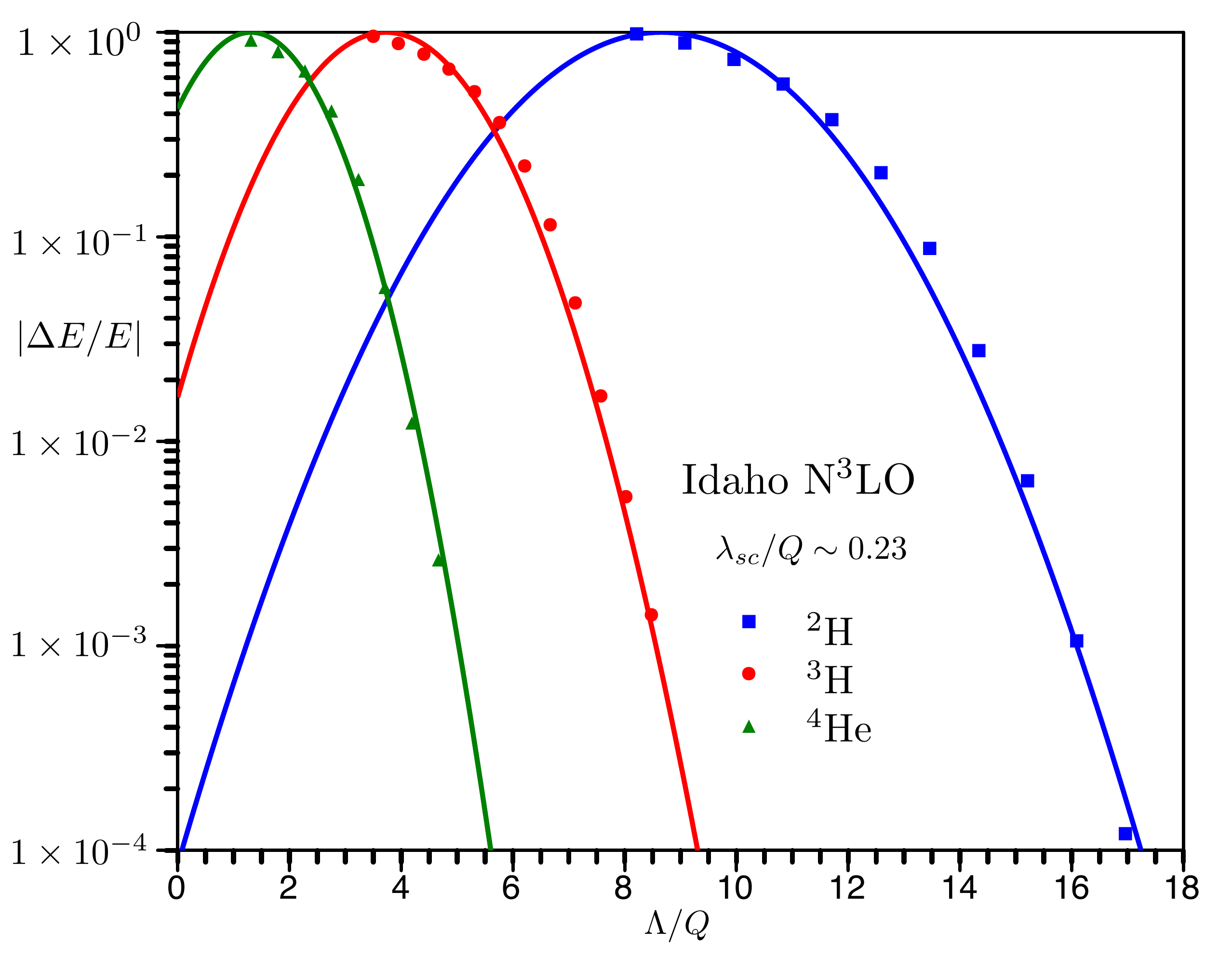}}
\caption{(Color online)  Dependence of the ground-state energy of three $s$-shell nuclei (compared to a converged value; see text) upon the uv  momentum cutoff    $\Lambda\leq \Lambda^{N3LO}$. The data are fit to Gaussians. Both uv and ir cutoffs are scaled to $Q$, the binding momentum of each nucleus, so that the $s$-state nuclei can be fit on a single plot.  The unscaled values are $\lambda_{sc} = 10$ MeV/$c$ for $^2$H, $\lambda_{sc} = 20$ MeV/$c$ for $^3$H and $\lambda_{sc} = 40$ MeV/$c$ for $^4$He.  }
\label{fig:12}
\end{figure}

Finally, we address an
 extrapolation in the uv cutoff toward the top of the shaded oval of Figure 5 to remove uv artifacts with the ir physics expected to be captured by binning or fixing values of $\lambda_{sc} < \lambda^{NN}_{sc}$.  To begin this exercise,  we return to Figure 4 and restrict our attention to the sector $\Lambda\leq\Lambda^{NN}$.  The universal curve in that sector  is generalized to  three $s$-shell nuclei  in Figure 12   (enhanced from Figure 9 of Ref. \refcite{Coon2012}) where all momenta are scaled by the binding momentum $Q$ of the considered nucleus in order to put them on the same plot.  The data in Figure 4  do not extend below $\Lambda\leq\sim 200$ MeV/$c$ but the lefthand side of the Gaussian fit to $\Lambda\leq\Lambda^{NN}$ is displayed in Figure 12 to guide the eye and to show that the peak of the Gaussian is not at $\Lambda=0$ MeV/$c$.

For such low fixed momenta $\lambda_{sc}$, $\vert\Delta E/E\vert$ does go to zero with increasing $\Lambda$ because $\lambda_{sc}\leq\lambda^{NN}_{sc}$. The  ``high" $\Lambda$  tails of these curves were fit by  Gaussians (shifted from the origin) in the variable $\Lambda/Q$ in Ref.~\refcite{Coon2012}.  This behavior suggests another possible extrapolation scheme; fixing the ir physics first and then extrapolating in the uv cutoff.  A later paper  did advocate  such a uv extrapolation with $\Lambda^2$ in the exponential function acting as the uv regulator \cite{FHP}.  We have tried to fit our data with the unshifted Gaussian ansatz, 
\begin{equation}
E_{gs}(\Lambda) =  A \exp{(-2\left({\Lambda/\lambda_{SRG}}\right)^2})+ E(\Lambda=\infty), 
 \label{eq:FHPuv}
\end{equation}
of that paper and failed,
 perhaps because  our calculations were made with the original Idaho N$^3$LO potential rather than the SRG-evolved interaction used in  Ref. \refcite{FHP}.  Because the Gaussians are shifted from the origin, a fit requires instead 
\[E_{gs}(\Lambda/Q) = a \exp(-{(\Lambda/Q-b)}^2/2c^2) + E(\Lambda/Q=\infty).\]
 Such extrapolations are shown in Figure 13 where the fit is restricted to the uv range  $\Lambda/Q\leq\Lambda^{NN}/Q$, as already shown in Figure 12.
 
\begin{figure}[htpb!]
\centerline{\includegraphics[width=0.90\textwidth]{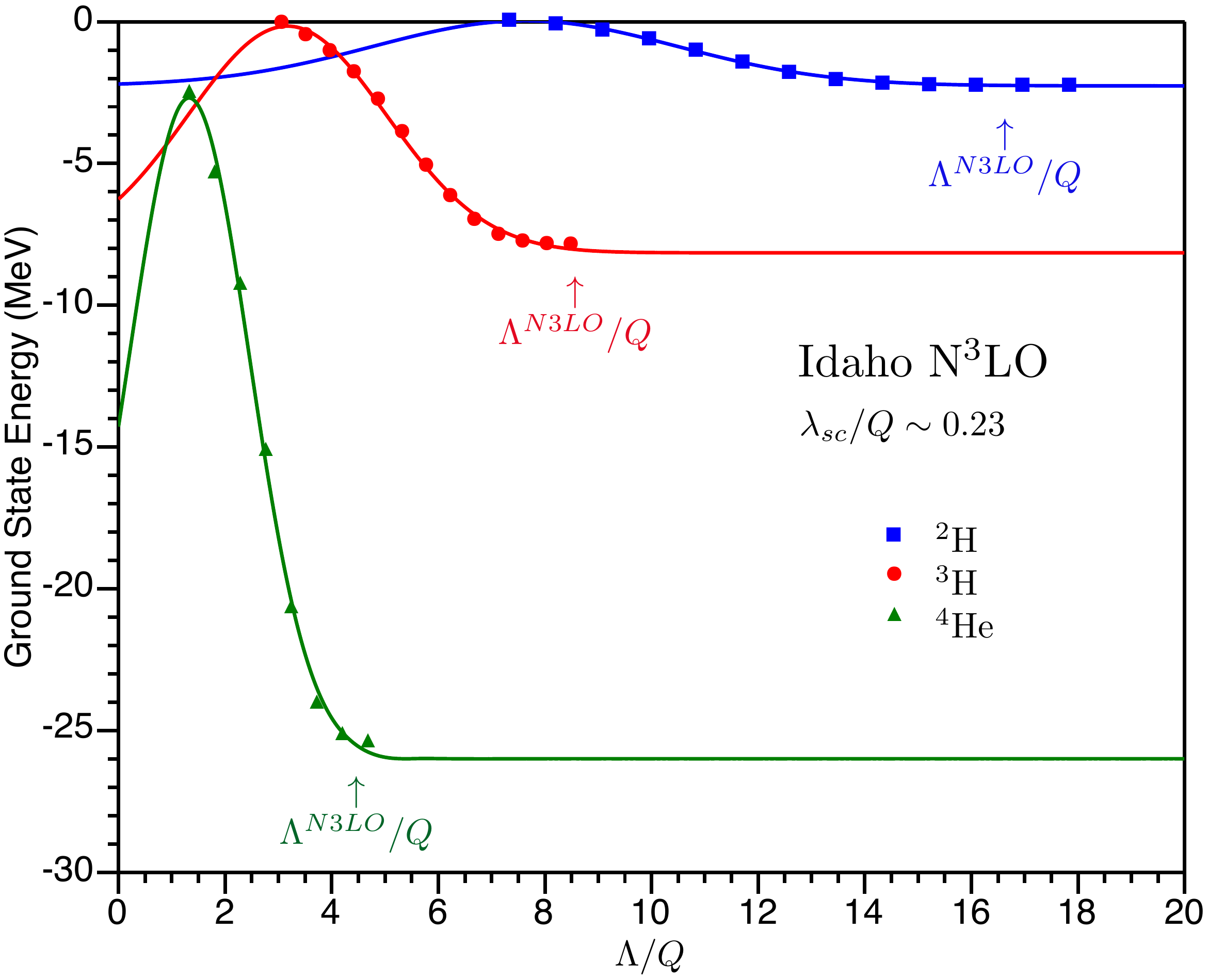}}
\caption{(Color online)  Extrapolation at fixed $\lambda_{sc}\leq \lambda_{sc}^{N3LO}$.  Both uv and ir cutoffs are scaled to $Q$, the binding momentum of each nucleus, so that the $s$-state nuclei can be fit on a single plot.  The unscaled values are $\lambda_{sc} = 10$ MeV/$c$ for $^2$H, $\lambda_{sc} = 20$ MeV/$c$ for $^3$H and $\lambda_{sc} = 40$ MeV/$c$ for $^4$He.  The arrows indicate that the UV extrapolation uses only points for which $\Lambda \leq \Lambda^{N3LO}$. }
\label{fig:13}
\end{figure}

Unfortunately, the extrapolated ground state energies of Figure 13 do not agree with those obtained from {our ir extrapolations or with independent calculations.  The extrapolated energies are always lower:  2 keV for the deuteron, 300 keV (or 4\%) for the triton and 620 keV(or 2.4\%) for the alpha particle.  It is difficult to achieve consistent extrapolations with different values of fixed (low) $\lambda_{sc}$.  For example, if one takes $\lambda_{sc} = 12$ MeV/$c$ instead of $\lambda_{sc} = 20$ MeV/$c$,  seemingly closer to the ir limit so that even more of the ir physics is captured, the extrapolated triton energy is -10.149 MeV;  2.3 MeV below the accepted value.  We did make a modest number of trials of a uv extrapolation of $^3$H  with an SRG transformed potential. Only with the SRG transformed potentials does the extrapolation procedure illustrated in Figure 13 agree with other independent calculations.

  A recent extensive study of uv extrapolations concludes that the phenomenological uv extrapolation formula as given by our Eq.~\ref{eq:FHPuv} (taken from Ref.~\refcite{FHP}) is an approximation; an approximation valid only in a limited range of 
$\Lambda/\lambda_{SRG}$.\cite{Koenig}  This study and derivations were made in the context of the two-body bound state.  A graphical demonstration of this limited range can be seen in Fig. 2b of 
Ref.~\refcite{Koenig} in which the dependence of $|\Delta E/E|$ upon the uv regulator  can be seen to be Gaussian only in the approximate range  $ 140- 460$~MeV/$c$ for the deuteron of the  Idaho
    N$^3$LO potential transformed by $\lambda_{SRG} = 2\, \rm{fm} ^{-1}$.  This figure is quite similar to Fig. 4; the starting point of our uv extrapolation exercise.  One finds also in Ref.~\refcite{Koenig} the demonstration in Figures 15a and 15b that a uv Gaussian extrapolation is worse, compared to the exact result, for the original Idaho  N$^3$LO potential than for the one softened by the SRG transformation.  We suggest that these findings are consistent with the results presented here.

It should be noted that a combined ir and uv extrapolation has been made for the ground state energy of $^4$He subject to the JISP16 potential without the Coulomb potential  and calculated with the  no-core Monte Carlo shell model (``single-particle-cut space")\cite{Abe_NTSE-2014}.  The preliminary result of  the combined extrapolation
\[E_{gs}(\lambda_{sc},\Lambda) = a \exp(-b/\lambda_{sc})  +
c \exp[-(\Lambda^2/d^2)] + E(\lambda_{sc}=0,\Lambda=\infty).\]
 agreed very well with the conventional ir extrapolation  Eq.~\ref{eq:irextrap} as $\lambda_{sc} \rightarrow 0$ and with the traditional extrapolation  Eq.~\ref{eq:traditional}  (which had larger errors as discussed in the Introduction.)       This potential is very soft and an extrapolation error analysis is forthcoming.

\section{Conclusions}

In conclusion, we have established that an extrapolation in the ir cutoff with the uv cutoff above the intrinsic uv scale of the interaction is quite successful, not only for the eigenstates of the Hamiltonian but also for  expectation values of operators considered long range.  On the other hand, an extrapolation in the uv cutoff when the ir cutoff is below the intrinsic ir scale is neither robust nor reliable.

\section{Acknowledgements}
Extremely useful conversations with Sean Fleming  are acknowledged.  We thank the authors of Ref. \refcite{Micah} for sharing with us their results before publication.   We also thank the Iowa State group for sharing with us their JISP16 results after publication \cite{Maris09,Cockrell12,Maris13}.
 We are grateful to  Petr Navr\'atil for generously allowing us to use his $\it manyeff$ Jacobi coordinate  code \cite{NKB} for some of our calculations.
 Numerical calculations have been performed in part at the LLNL LC facilities supported
by LLNL under Contract No. DE-AC52-07NA27344. This contribution was supported in part by USDOE Division of Nuclear Physics grant DE-FG02-04ER41338 (Effective Theories of the Strong Interaction).

\end{document}